\documentclass[twocolumn,showpacs,preprintnumbers,amsmath,amssymb,nofootinbib]{revtex4}
\usepackage{dcolumn}
\usepackage{bm}
\usepackage{tabularx}
\usepackage{array}
\usepackage{graphics}
\usepackage{graphicx}
\usepackage{psfrag}
\usepackage{epsfig}
\usepackage{amsmath}
\usepackage{amssymb}
\usepackage{multirow}
\usepackage{verbatim}
\usepackage{color}
\usepackage{psfrag}
\usepackage[usenames,dvipsnames]{xcolor}

\usepackage{verbatim}

\newcommand{\msbar}{{\overline{\rm MS}}}

\newcommand{\nf}{N_{\rm f}}
\newcommand{\bea}{\begin{eqnarray}}
\newcommand{\eea}{\end{eqnarray}}
\newcommand{\bel}{\begin{align}}
\newcommand{\eel}{\end{align}}
\newcommand{\beq}{\begin{equation}}
\newcommand{\eeq}{\end{equation}}
\newcommand{\gev}{{\rm GeV}}

\newcommand{\pdir}{p\kern -5.2pt\raise 0.2ex\hbox {/}}
\newcommand{\vdir}{v\kern -5.75pt\raise 0.15ex\hbox {/}}
\newcommand{\kdir}{k\kern -5.75pt\raise 0.15ex\hbox {/}}
\newcommand{\epsdir}{\epsilon\kern -5.0pt\raise 0.15ex\hbox {/}}
\newcommand{\bvdir}{\bar{v}\kern -5.75pt\raise 0.15ex\hbox {/}}
\newcommand{\Ddir}{D\kern -7.75pt\raise 0.20ex\hbox {/}}
\newcommand{\Adir}{A\kern -7.75pt\raise 0.20ex\hbox {/}}
\newcommand{\ldir}{l\kern -5.0pt\raise 0.2ex\hbox{/}}
\newcommand{\varepsdir}{\varepsilon\kern -5.5pt\raise 0.15ex\hbox{/}}

\newcommand{\gone}{{\cal G}(1)}

\newcommand{\gw}{{\cal G}(w)}

\def\l{\left}
\def\r{\right}
\def\ds{\displaystyle}
\newcommand{\nn}{\nonumber}
\makeatother

\begin{document}
\preprint{\tt LPT 13-74}
\vspace*{22mm}
\title{$B_s\to D_s\ell \nu_\ell$ near zero recoil in and beyond the Standard Model}

\author{Mariam Atoui}
 \email{matoui@in2p3.fr}
\affiliation{%
Laboratoire de Physique Corpusculaire, Universit\'e Blaise Pascal,  
CNRS/IN2P3, 63177 Aubi\`ere Cedex, France}

\author{Damir Be\v{c}irevi\'c}
 \email{Damir.Becirevic@th.u-psud.fr}
\affiliation{%
Laboratoire de Physique Th\'eorique (B\^at 210), Universit\'e Paris Sud and CNRS,  
Centre d'Orsay, 91405 Orsay-Cedex, France}

\author{Vincent Mor\'enas}
 \email{morenas@in2p3.fr}
\affiliation{%
Laboratoire de Physique Corpusculaire, Universit\'e Blaise Pascal,  
CNRS/IN2P3, 63177 Aubi\`ere Cedex, France}

\author{Francesco Sanfilippo} 
\email{Francesco.Sanfilippo@th.u-psud.fr}
\affiliation{%
Laboratoire de Physique Th\'eorique (B\^at 210), Universit\'e Paris Sud and CNRS,  
Centre d'Orsay, 91405 Orsay-Cedex, France}

\date{\today}

\begin{abstract}
We compute the normalization of the form factor entering the $B_{s}\to D_{s}\ell \nu$ decay amplitude by using numerical simulations of QCD on the lattice. 
From our study with $\nf=2$ dynamical light quarks, and by employing the maximally twisted Wilson quark action, we obtain in the continuum limit $\gone = 1.052(46)$.  
We also compute the scalar and tensor form factors in the region near zero recoil and find $f_0(q_0^2)/f_+(q_0^2)=0.77(2)$, $f_T(q_0^2,m_b)/f_+(q_0^2)=1.08(7)$, for $q_0^2=11.5\ \gev^2$.
These latter results are useful for searching the effects of physics beyond the Standard Model in $B_{s}\to D_{s}\ell \nu$ decays. Our results for the similar form factors relevant 
to the non-strange case indicate that the method employed here can be used to achieve the precision determination of the $B\to D\ell \nu$ decay amplitude as well. 
\end{abstract}

\maketitle

\section{\label{Introduction}Introduction}
Inclusive and exclusive semileptonic $b\to c \ell \nu$ decays, with $\ell \in\{e,\mu \}$, have been subjects of intensive research over the past two decades. Within the Standard Model (SM) the main target of that research was, and still is, the accurate determination of the Cabibbo--Kobayashi-Maskawa matrix element $|V_{cb}|$, which is extracted from the comparison of theoretical expressions  with experimental measurements of the partial or total decay widths.  It turns out, however, that the value for $|V_{cb}|$ obtained from the exclusive decays agrees only at the $2.1\ \sigma$ level with the one extracted from inclusive decays. More specifically, the two independent analyses of inclusive decays~\cite{kinetic,1S} (updated in refs.~\cite{PDG}) give completely consistent results which, after taking an average, reads~\footnote{The most recent update of the analysis of inclusive decays using the kinetic scheme can be found in ref.~\cite{gambino}.}
\bea\label{Vcbincl}
\vert V_{cb}\vert_{\rm incl.} = 41.90(70)\cdot 10^{-3}.
\eea
The analyses of exclusive decays, instead, are performed by fitting the experimental data to the shapes of the form factors parameterized according to the expressions proposed and derived in ref.~\cite{CLN}, so that the final results are then reported in the following form,  
\bea
&& \vert V_{cb}\vert {\cal F}(1) = 35.90(45) \cdot 10^{-3}\,,\nn\\
&&    \nn\\
&& \vert V_{cb}\vert {\cal G}(1) = 42.6(1.5) \cdot 10^{-3}\,,
\eea
as obtained from $B(B\to D^\ast \mu \nu)$ and $B(B\to D \mu \nu)$, respectively. ${\cal F}(1)$ and ${\cal G}(1)$ are the relevant hadronic form factors at the zero recoil point. Thanks to the heavy quark symmetry, and up to perturbative QCD corrections, both these form factors are equal to one in the limit of $m_{c,b}\to \infty$~\cite{isgurwise}. To compute the deviation of these form factors from unity, it is necessary to include all the non-perturbative order $1/m_{c,b}^n$ QCD corrections. The only model independent method allowing to compute ${\cal F}(1)$ and ${\cal G}(1)$ from the first theory principles is lattice QCD. Using the most recent estimates of the above form factors, ${\cal F}(1)= 0.902(17)$~\cite{bailey2010} and ${\cal G}(1)=1.074(24)$~\cite{Okamoto:2004xg}, one arrives to 
\bea
\vert V_{cb}\vert_{\rm excl.} = 38.56(89) \cdot 10^{-3},
\eea
a number that obviously differs from $\vert V_{cb}\vert_{\rm incl.}$ given in eq.~(\ref{Vcbincl}).
In principle the exclusive decay modes are better suited for the precision determination of $\vert V_{cb}\vert$ because less theoretical assumptions are needed to compute the corresponding decay rates.  The main obstacle is, however, the necessity for a reliable, high precision, lattice QCD estimate of  ${\cal F}(1)$ and ${\cal G}(1)$. Furthermore, for the required percent precision of $\vert V_{cb}\vert$ it is important to have a good control over the structure dependent soft photon $B\to D^{(\ast)}\ell \nu \gamma_{\rm soft}$ which could otherwise be misidentified as pure semileptonic decays. This problem is less acute for the down-type spectator, i.e.  $\bar B^0\to D^{(\ast) +}\ell \bar \nu$, than for the up-type spectator quark, $B^-\to\overline D^{(\ast) 0}\ell \bar \nu$~\cite{kosnik}. For that reason it is desirable to consider the charged and neutral $B$-meson decay modes separately. In this paper we will mainly discuss $B_s\to D_s\ell \nu$ decay for which the soft photon pollution is smaller. Moreover, this mode is much more affordable numerically because the valence $s$-quark is easily accessible in numerical simulations of QCD on the lattice which is not the case with the physical $u/d$-quark.  
In this paper we will also comment on the non-strange case when appropriate. Finally, we prefer to focus on  $B_{(s)}\to D_{(s)}\ell \nu$, rather than  $B_{(s)}\to D_{(s)}^\ast \ell \nu$, because the hadronic matrix element involves much less form factors and the decay rate is therefore likely to be less prone to systematic uncertainties.

Despite the importance of $\gone$, only a few lattice QCD studies have been performed so far. The methodology described in ref.~\cite{hashimoto} has been implemented in unquenched simulations with $N_f=2+1$  dynamical staggered light quarks in refs.~\cite{Okamoto:2004xg,Bailey:2012rr} where the propagating heavy quarks on the lattice have been interpreted by means of an effective theory approach. In ref.~\cite{nazarioA}, an alternative method to compute the $B\to D$ semileptonic form factors has been proposed and implemented in quenched approximation. That latter method, based on use of the step scaling function,  allows to compute the same form factors without  recourse to heavy quark effective theory. The price to pay, however, is that the method of ref.~\cite{nazarioA} is computationally very costly and to this date it has not been extended to unquenched QCD. In this paper we use a modification of the proposal of ref.~\cite{nazarioA}, presented in ref.~\cite{blossier} and also implemented in the computation of the decay constant $f_B$ and the $b$-quark mass, cf. ref.~\cite{dimopoulos}. The remainder of this paper is organized as follows: In sec.~\ref{sec:Definitions} we define the form factors and express them in a way that is suitable for the strategy used for their computation which is described in sec.~\ref{sec:strategy}. Details of our lattice computations and the results for $\gone$ are given in sec.~\ref{sec:Lattices}, while the results concerning the scalar and tensor form factors in the region close to zero recoil are discussed and presented in sec.~\ref{sec:bsm}. We finally conclude in sec.~\ref{sec:conclusions}.

\section{Definitions\label{sec:Definitions}}
The hadronic matrix element describing the $B_s\to D_s\ell \nu_\ell$ decay in the SM, $\langle D_s\vert \bar b \gamma_\mu (1-\gamma_5)c \vert B_s\rangle$ $\equiv \langle D_s\vert \bar b \gamma_\mu c \vert B_s\rangle$, is parameterized in terms of the hadronic form factors $f_{+,0}(q^2)$ as
\begin{align}\label{eq:def1}
\langle D_s(k)\vert V_\mu\vert & B_s(p)\rangle = \l( p + k \r)_\mu f_+(q^2) + \nn\\
&q_\mu {m_{B_s}^2-m_{D_s}^2\over q^2} \l[ f_0(q^2) - f_+(q^2)\r]\,,
\end{align}
where $V_\mu = \bar b \gamma_\mu c$, $q = p-k$, and $q^2\in (0, q^2_{\rm max}]$, with $q^2_{\rm max}=(m_{B_s}-m_{D_s})^2$. 
The extraction of the form factors becomes particularly simple if we use the projectors
\bea\label{proj}
\mathbb{P}_\mu^0 = {q_\mu\over q^2_{\rm max}}\,,\qquad \mathbb{P}_\mu^+ = \left({{\vec q\ }^2\over m_{B_s}-E_{D_s}},\vec q \right)\,,
\eea
so that
\begin{align}\label{eq:ffs}
\mathbb{P}_\mu^0 \langle D_s(k)\vert &V_\mu\vert {B_s}(p)\rangle = {m_{B_s}+m_{D_s}\over m_{B_s}-m_{D_s}} f_0(q^2)\,,\nn\\
\mathbb{P}_\mu^+ \langle D_s(k)\vert &V_\mu\vert B_s(p)\rangle = {\vec q\ }^2 {2 m_{B_s}\over m_{B_s} - E_{D_s}} \ f_+(q^2)\,.
\end{align}
In our computations we will consider the situations with $|\vec p |=0$ and $q_\mu=(m_{B_s}-E_{D_s}, -\vec k)$. Another frequently used parameterization 
of this matrix element, motivated by the heavy quark expansion, reads 
\begin{align} \label{eq:def2}
{1\over \sqrt{m_{B_s}m_{D_s}}} \langle {D_s}(k)\vert V_\mu\vert {B_s}(p)\rangle 
&= \l( v + v^\prime \r)_\mu h_+(w) \nn\\ 
&+ \l( v - v^\prime \r)_\mu h_-(w)\,,
\end{align}
where $v=p/m_{B_s}$, $v^\prime=k/m_{D_s}$, and the relative velocity $w= v\cdot v^\prime=(m_{B_s}^2+m_{D_s}^2-q^2)/(2 m_{B_s} m_{D_s})$.  
From the comparison of eqs.~(\ref{eq:def1}) and~(\ref{eq:def2}) one gets
\begin{align}\label{eq:f2h}
f_+(q^2)=
{m_{B_s}+m_{D_s}\over \sqrt{4 m_{B_s}m_{D_s}}} & h_+(w) 
\left[ 1 - {m_{B_s}-m_D\over m_{B_s}+m_{D_s}} {h_-(w)\over h_+(w)}\right]\,,  \nn\\
 f_0(q^2)={\sqrt{ m_{B_s} m_{D_s}}\over m_{B_s}+m_{D_s}}& \l( w+1\r) h_+(w) \biggl[ 1 - \biggr.  \nn\\
&\biggl. \qquad {m_{B_s}+m_{D_s}\over m_{B_s}-m_{D_s}} {w-1\over w+1}{h_-(w)\over h_+(w)}\biggr]\,.
\end{align}
The form factor $\gw$ used in experimental analyses of  the $B_{(s)} \to D_{(s)}\ell\nu$ decay is proportional to $f_+(q^2)$, and reads
\bea\label{eq:Hw2}
\gw &=& h_+(w) \left[ 1 - {m_{B_s}-m_{D_s}\over m_{B_s}+m_{D_s}} {h_-(w)\over h_+(w)}\right] \nn\\
&\equiv & h_+(w) \left[ 1 - \left({m_{B_s}-m_{D_s}\over m_{B_s}+m_{D_s}}\right)^2 H(w) \right],
\eea
where, for convenience, we introduced  
\bea
H(w) =  {m_{B_s}+m_{D_s}\over m_{B_s}-m_{D_s}} {h_-(w)\over h_+(w)}\,.
\eea
Our main target is the determination of ${\cal G}(1)$, and therefore we are particularly interested in the dominant $h_+(1)$ term that can be easily obtained from $f_0(q^2_{\rm max})$, 
\bea\label{eq:step1}
h_+(1) = {   m_{B_s}+m_{D_s} \over  \sqrt{4 m_{B_s} m_{D_s}} } f_0(q^2_{\rm max}) \,.
\eea
Unfortunately, however, $H(1)$ is not directly accessible from the lattice. 
Instead, we need to compute the form factors $f_{0,+}(q^2)$ at several small values of the $D$-meson three-momentum and then extrapolate $H(w)$ to $H(1)$. 
As we shall see in the following, we manage to work with $w\gtrsim 1$ but by staying very close to zero recoil and the uncertainty associated with this extrapolation is completely negligible. 
To be more specific, at $|\vec k |\neq 0$, we compute 
\bea
R_0(q^2)={f_0(q^2)\over f_+(q^2)}\,,
\eea
so that 
\begin{align}
H(w)= { R_0 ( m_{B_s}+m_{D_s})^2 - 2  m_{B_s} m_{D_s} (w +1)\over R_0 ( m_{B_s}-m_{D_s})^2 - 2  m_{B_s} m_{D_s} (w -1) }\,,
\end{align}
where, for shortness, we write $R_0\equiv R_0(q^2(w))$.

\section{Strategy\label{sec:strategy}}

To extract the form factors $f_{0,+}(q^2)$ from numerical simulations on the lattice, one first  computes the correlation functions
\begin{align}\label{eq:00}
C_{\mu}(\vec q;t)= \sum_{\vec x,\vec y } \langle P_{bs}(\vec  0,0) V_\mu (\vec x, t)   P_{cs}^\dag (\vec y, t_S) e^{-i\vec q (\vec x-\vec y)}\rangle\,, 
\end{align}
where the interpolating source operators, the pseudoscalar densities $P_{cs}$ and $P_{bs}$, are sufficiently separated in the time direction so that for $0\ll t\ll t_S$ one can isolate the lowest lying states with $J^P=0^-$ that couple to two source operators, and then extract the matrix element of the vector current between the two. The simplest choice would be the local operator, $P_{hs}=\bar h\gamma_5 s$, but for practical convenience one often resorts to the smearing technique that helps to significantly reduce the couplings to radially excited states. In other words,  the lowest lying states are better isolated when smeared source operators are used and when the corresponding time interval, within which the matrix elements are extracted, becomes larger. As mentioned in the previous section, we also need to give the $D_s$ meson a few momenta $|\vec k |\neq 0$, in order to study the behavior of $H(w)$ as a function of $w$ and extrapolate to $H(1)$. To make those momenta small and remain close to the zero recoil point, it is convenient to compute the quark propagators, $S_q(x,0;U) \equiv \langle q(x)\bar q(0)\rangle$,  by imposing the twisted boundary conditions~\cite{nazario}. Those are easily implemented by rephasing the gauge field configurations according to,  
\bea\label{twisted}
U_\mu(x) \to U^\theta_\mu(x) = e^{i \theta_\mu \pi/L}U_\mu(x)\ ,
\eea
where  $U_\mu(x)$ stands for the gauge links, $ \theta_\mu = (0,\vec \theta)$, and $L$ is the size of the spatial side of the cubic lattice. The quark propagator computed on such a rephased configuration, 
\bea\label{eq:tbc}
S_q^{ \vec \theta} (x,0;U)  = e^{i \vec \theta\cdot \vec x \pi/L} \ S_q(x,0;U^\theta) \,,
\eea
can then be combined with the ordinary (untwisted) propagator, $S_q(x,0;U)$, into a two-point correlation function. The resulting lowest lying state extracted from the exponential fall-off has a three-momentum different from zero,  $\vert \vec k\vert = \vert \vec \theta \vert \pi/L$~\cite{nazario,chris-giovanni,diego}. Then, by choosing $\vec \theta = (1,1,1) \times \theta_0$, which also minimizes the discretization errors, one can tune $\theta_0$ to an arbitrary small value and therefore explore the kinematical region of $B_s\to D_s\ell \nu_\ell$ decay very close to zero recoil, $w\gtrsim 1$ ($q^2\lesssim q^2_{\rm max}$).

For a given $\theta_0$, the continuum energy-momentum relation would give
\bea\label{eq:ww}
w=\sqrt{1 + {3 \theta_0^2\pi^2\over m_{D_s}^2 L^2}}\,,
\eea
which is a good approximation for the small values of $\theta_0$ chosen in this study. Otherwise one can use $w=E(\theta_0)/m_{D_s}$, and the free boson dispersion relation on the lattice:
\begin{align}
4\sinh^2{E_D\over 2} = 4 \left( 3 \sin^2{\theta_0\pi\over 2 L}\right) +   4 \sinh^2{m_{D}\over 2}\,.\nn
\end{align}
In terms of quark propagators the correlation function~(\ref{eq:00}) reads
 \begin{equation}\label{eq:three-tw}
\begin{split} 
&C_{\mu}(\vec q;t) = \\
& \langle \sum_{\vec x,\vec y}{\rm Tr}\left[  \gamma_5 S_s(0,y) \gamma_5 S^{ \vec \theta}_c (y,x;U)\gamma_\mu S_b(x,0;U)  \right]\rangle \,.
\end{split}
\end{equation}
In addition to the above three-pont correlation functions, we also computed the two-point correlators that are necessary to remove the source operators and gain access to the vector current matrix element. 
From  the large time behavior of the two-point  correlation functions
\begin{equation} \label{eq:2pts}
\begin{split} 
 \langle {\displaystyle \sum_{\vec x} }  P_{hs}&(\vec x; t)  P_{hs}^\dagger (0; 0) \rangle  \xrightarrow[]{\displaystyle{ t\gg 0}}  \\
 &\;  \left| {\cal Z}_H \right|^2 \frac{\cosh[  m_H (T/2-t)]}{ m_H } e^{- m_H T/2}\,,
\end{split}
\end{equation}
we can extract $m_H$ and ${\cal Z}_{H}=\langle 0| \bar h\gamma_5 s|H\rangle$, where  $h$ ($H$) stands for either $c$ ($D_s$) or $b$ ($B_s$), and  $T$ is the size of the temporal extension of the lattice. 
With these ingredients we are able to extract the desired hadronic matrix element from decomposition
\begin{equation} \label{eq:3pts}
\begin{split} 
C_{\mu}(\vec q;t)  &  \xrightarrow[]{\displaystyle{0 \ll  t\ll t_S}}  {  {\cal Z}_{B_s}{\cal Z}_{D_s} \over 4 m_{B_s} E_{D_s}}  \exp\l( -m_{B_s}t \r)\times \\
 &\;   \exp\l[-E_{D_s}( t_S-t) \r]  \langle D_s (\vec k) \vert V_\mu \vert B_s (\vec 0)\rangle 
 \,,
\end{split}
\end{equation}
and then by using the projectors (\ref{proj}) we get the form factors $f_{0,+}(q^2)$, as indicated in eq.~(\ref{eq:ffs}).
As already mentioned above, to make sure that the lowest lying states are well isolated, we employed the smearing procedure discussed in our previous works~\cite{charm2} where the full information concerning the smearing parameters can be found as well.

While working with the fully propagating $b$-quark on the lattice, i.e. without resorting  to an effective theory approach, the most difficult problem is to deal with large discretization errors. The reason is mostly practical since the lattice spacings ($a$) accessible in current lattice QCD simulations are not small enough to satisfy $m_b a < 1$. The charm quark, on the other hand, can be simulated directly and the discretization effects associated with its mass can be monitored by working at several small lattice spacings. Hence, the strategy is to perform the computations starting from the charm quark mass and then successively increase the heavy quark mass by a factor of $\lambda$ so that after $n+1$ steps one arrives at $m_b$. For each value of the heavy quark mass $m_h=\lambda^{k+1} m_c$ we compute the form factor ${\cal G}(1,m_h,m_c)$, and evaluate the ratio of form factors computed at two successive heavy quark masses, while keeping other valence quarks and the lattice spacing fixed. In practice we compute 
\bea\label{eq:bigSigma}
\Sigma_k(1) = {  {\cal G}(1,\lambda^{k+1} m_c,m_c, a^2) \over {\cal G}(1,\lambda^k m_c,m_c,a^2) }\,,
\eea
where the first argument in the form factor is $w=1$, the second is the heavy quark mass that we want to send to the physical $b$-quark, while $m_c$ and $a$ are the charm quark mass and the lattice spacing, both of which are kept fixed. 
Each of these ratios can then be extrapolated to the continuum limit, ${\ds{\lim_{a\to 0}}} \Sigma_k(1)= \sigma_k(1)$. 

The advantage of considering $\sigma_k(1)$ instead of ${\cal G}(1,\lambda^{k+1}m_c,m_c)$ becomes apparent when considering the heavy quark mass dependence. 
In the continuum limit, thanks to the heavy quark symmetry, the form factor scales with the inverse heavy quark mass as 
\bea\label{eq:hqe1}
{\cal G}(1,m_h,m_c) = g_0 + \frac{g_1}{m_h} + \frac{g_2}{m_h^2} + \dots , 
\eea
where the non-perturbative coefficients $g_{0,1,2,\dots}$ should be determined from the fit to lattice data. Keeping in mind the practical limitations that prevent us to ensure that the heavy quark masses are smaller than the inverse lattice spacing, it is clear that it is very challenging to disentangle the physical effects from lattice artifacts in $g_i$'s, and in the dominant term $g_{0}$ in particular. Consequently the resulting $\gone\equiv {\cal G}(1,m_b,m_c)$ suffers from systematic uncertainty, the size of which is very difficult (if not impossible) to assess and therefore cannot be used for a precision determination of $\vert V_{cb}\vert$ from $B(B_{(s)}\to D_{(s)}\ell \nu)$. 
In contrast, the successive ratios of the form factors satisfy ${\ds{\lim_{m_h\to \infty}}} \sigma(m_h) = 1$, and therefore instead of extrapolating 
to the inverse $b$-quark mass, one actually interpolates to $\sigma(m_b)$. In the continuum limit, we then fit the lattice data to
\bea\label{form:2}
\sigma(1,m_h,m_c) \equiv \sigma(1,m_h) = 1 + \frac{s_1}{m_h} + \frac{s_2}{m_h^2} + \dots \ ,
\eea
determine $s_{1,2}$, and interpolate to $\sigma(m_b)$.  Another interesting feature is that the expansion in inverse heavy quark mass is strictly valid in the heavy quark effective theory (HQET) and had we used eq.~(\ref{eq:hqe1}) we would have had to include the perturbative matching between our results (obtained in full QCD) to HQET, and then the result of extrapolation to the $b$ quark mass should have been converted back to QCD. In the ratios of form factors, $\Sigma_k$ ($\sigma_k$), the matching to HQET and back becomes completely immaterial as the matching factors cancel to a large extent. We attempted including these corrections to our interpolation to $\sigma(1,m_b)$, and the results remained would change by a few per-mil level only, thus completely immaterial for our purpose.

To get the physically relevant  $\gone$, one starts from the elastic form factor,  the value of which is by definition $ {\cal G}(1, m_c,m_c) = 1$. 
The physically interesting $B_{(s)}\to D_{(s)}$ form factor is then obtained as a product of $\sigma_k(1)$ factors discussed above, namely
\bea\label{eq:main1}
\gone&\equiv & {\cal G}(1, m_b,m_c) \nn\\
&=& \sigma_n \sigma_{n-1}\dots \sigma_1 \sigma_0 \underbrace{ {\cal G}(1, m_c,m_c) }_{=1}\,. 
\eea
In this study we choose $n=8$, which gives
\bea
\lambda = \l( \frac{m_b}{m_c}\r)^{\frac{1}{n+1}} = 1.176\,,
\eea
where we used $m_c^\msbar (2\ \gev)=1.14(4)$~GeV  \cite{charm1}, and $m_b^\msbar (2\ \gev)=4.91(15)$~GeV \cite{dimopoulos}. 

\section{Lattice details\label{sec:Lattices}}

In this work we use the publicly available gauge field configurations that include $\nf=2$ dynamical light quarks, generated according to the twisted mass QCD action with maximal twist~\cite{fr} by the European Twisted Mass Collaboration (ETMC)~\cite{Boucaud:2008xu}. 
Main details about $13$ ensembles of gauge field configurations are collected in tab.~\ref{tab:01}. 
We computed all quark propagators by using stochastic sources, and then applied  the so-called one-end trick to compute the correlation functions~\cite{Boucaud:2008xu}.

\begin{figure}[h!]
\centering
\includegraphics[scale=.56]{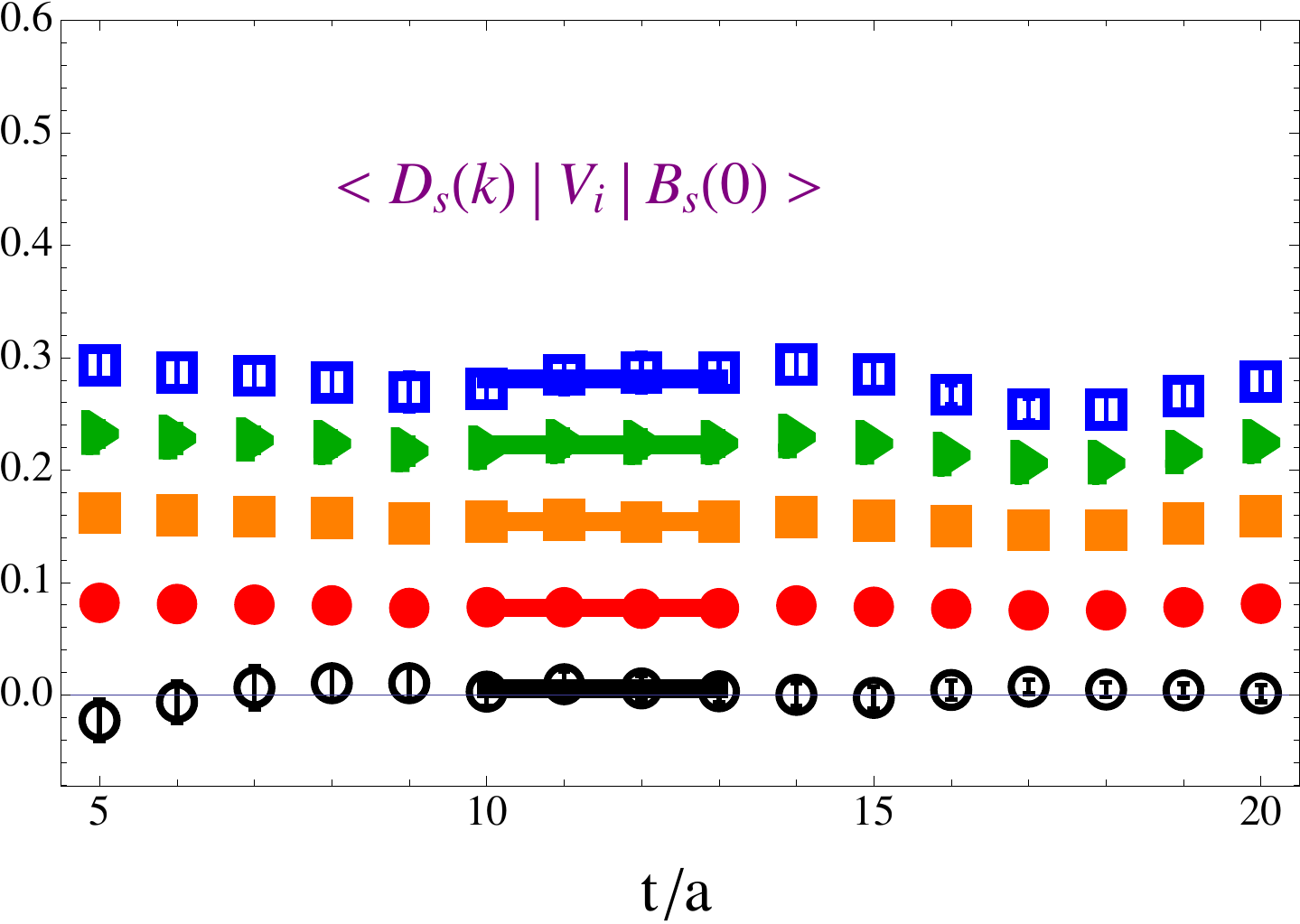}
\caption{\label{fig:plateaus} \sl Plateaus on which the matrix element is extracted according to eq.~(\ref{eq:3pts}) for $5$ different values of momenta corresponding to $w=1$ (when this matrix element is zero) and four other momenta corresponding to $w$'s given in eq.~(\ref{nonzero}). Plotted are the data from the ensemble IV (cf. tab.~\ref{tab:01})  and for $m_b=m_h=\lambda^4 m_c$ (N.B. $m_b^{\rm phys.}=\lambda^9 m_c$).}
\end{figure}

\renewcommand{\arraystretch}{2} 
\begin{table*}[h!!]
\begin{ruledtabular}
\begin{tabular}{|c|ccccccccc|}
Ensemble & $\beta =6/g_0^2$ & $\mu_{{\rm sea}}$ & $L^{3}\times T$ & $\#\,{\rm meas.}$ & $\mu_{s}$ & $\mu_{c}$ & $a\,{\rm [fm}]$ & $Z_{V}(g_{0}^{2})$ & $\ensuremath{Z_{T}(g_{0}^{2})}$\\
\hline 
I & \multirow{2}{*}{3.80} & 0.0110 & \multirow{2}{*}{$24^{3}\times48$} & 240 & \multirow{2}{*}{0.0194} & \multirow{2}{*}{0.2331 } & \multirow{2}{*}{0.098(3)} & \multirow{2}{*}{0.5816(2)} & \multirow{2}{*}{0.73(2)}\\
II &  & 0.0080 &  & 240 &  &  &  &  & \\
\hline 
III & \multirow{6}{*}{3.90} & 0.0100 & \multirow{4}{*}{$24^{3}\times48$} & 240 & \multirow{6}{*}{0.0177} & \multirow{6}{*}{0.2150} & \multirow{6}{*}{0.085(3)} & \multirow{6}{*}{0.6103(3)} & \multirow{6}{*}{0.750(9)}\\
IV &  & 0.0085 &  & 240 &  &  &  &  & \\
V &  & 0.0064 &  & 240 &  &  &  &  & \\
VI &  & 0.0040 &  & 288 &  &  &  &  & \\
VII &  & 0.0040 & \multirow{2}{*}{$32^{3}\times64$} & 240 &  &  &  &  & \\
VIII &  & 0.0030 &  & 240 &  &  &  &  & \\
\hline 
IX & \multirow{3}{*}{4.05} & 0.0080 & \multirow{3}{*}{$32^{3}\times64$} & 686 & \multirow{3}{*}{0.0154} & \multirow{3}{*}{0.1849} & \multirow{3}{*}{0.067(2)} & \multirow{3}{*}{0.6451(3)} & \multirow{3}{*}{0.798(7)}\\
X &  & 0.0060 &  & 400 &  &  &  &  & \\
XI &  & 0.0030 &  & 750 &  &  &  &  & \\
\hline 
XII & \multirow{2}{*}{4.20} & 0.0065 & $32^{3}\times64$ & 480 & \multirow{2}{*}{0.0129} & \multirow{2}{*}{0.1566} & \multirow{2}{*}{0.054(1)} & \multirow{2}{*}{0.686(1)} & \multirow{2}{*}{0.822(4)}\\
XIII &  & 0.0020 & $48^{3}\times96$ & 100 &  &  &  &  & \\
\end{tabular} 
{\caption{\footnotesize  \label{tab:01} Ensembles of gauge field condigurations used in this work. Lattice spacing is fixed by using the parameter $r_0/a$~\cite{R0}, with $r_0=0.440(12)$~fm fixed by matching $f_\pi$ obtained on the same lattices with its physical value  (cf. ref.~\cite{charm1}). Bare quark masses $\mu_i$ are given in lattice units. Quoted values of the renormalization constant $Z_T$ refer to the $\msbar$ renormalization scheme and $\mu = 2\ \gev$.}}
\end{ruledtabular}
\end{table*}
Since we use the smeared source operators, the factor $|{\cal Z}_{D_s}|$ needed in eq.~(\ref{eq:3pts}) depends on the momentum $\vec k$ given to the $D_s$ meson. We computed $|{\cal Z}_{D_s}(\vec k)|$ for each of our $\theta_0$-values and after dividing out the correlation function $C_\mu(t,\vec k)$ in eq.~(\ref{eq:three-tw}), by the exponentials and couplings $|{\cal Z}_{D_s}(\vec k)|$  and $|{\cal Z}_{B_s}|$, we looked for the plateau region to extract the desired matrix element [cf. eq.~(\ref{eq:3pts})]. 
After inspection, we fixed the plateaus to the intervals 
\begin{align} 
t_\beta \in [10,13]_{3.8}, [10,13]_{3.9}, [12,17]_{4.05},[17,19]_{4.2}, 
\end{align}
in an obvious notation.  Those plateaus are chosen to be common to all the sea quark masses considered at a given lattice spacing, and to all the heavy valence quark masses.  Note that the three-point correlation functions~(\ref{eq:00}) are computed with $t_S=T/2$ for all of our lattices. As an example, we illustrate in  fig.~\ref{fig:plateaus} the quality of the plateaus corresponding to the matrix element $\langle D_s(\vec k)|V_i|B_s(\vec 0)\rangle$, and their sensitivity to the values of the three-momentum $\vec k$ used in this paper  [or better, to the values of $\theta_0$ in (\ref{eq:three-tw})]. 
In tab.~\ref{tab:01} we gave the value of the charm quark mass in lattice units. Other heavy quark masses are simply obtained after successive multiplication by $\lambda$, with the physical $\mu_b=\lambda^9 \mu_c$. We note that the errors on the form factors become large for very heavy quarks. 

As mentioned in sec.~\ref{sec:strategy} the computation of $H(w)$ can be made only at $w\neq 1$. We tuned the values of the twisting angle $\theta_0$ for each of our lattices in such a way as to make the corresponding $w$ fixed [cf. eq.~(\ref{eq:ww})]. More specifically, apart from the zero recoil point $w=1$, we computed the form factors with four different non-zero momenta given to $D_s$, corresponding to 
\bea\label{nonzero}
w\in \{ 1.004, 1.016, 1.036, 1.062\}\,. 
\eea
Clearly, only a tiny extrapolation $H(w)$ to $H(1)$ is needed. Both a linear and a quadratic fit in $w$ to reach $H(1)$ lead to indistinguishable results, both results being small and further suppressed by the mass factor in eq.~(\ref{eq:Hw2}) so that $\gone$ is largely dominated by the form factor $h_+(1)$ evaluated according to eq.~(\ref{eq:step1}). For very heavy quarks ($m_h$ close to $m_b$) the effect of that extrapolation becomes visible, but since the form factor computed with such a heavy quark is dominated by discretization and large statistical errors they do not have any significant effect on our final results.

\begin{figure}[h!]
\centering
\includegraphics[scale=.65]{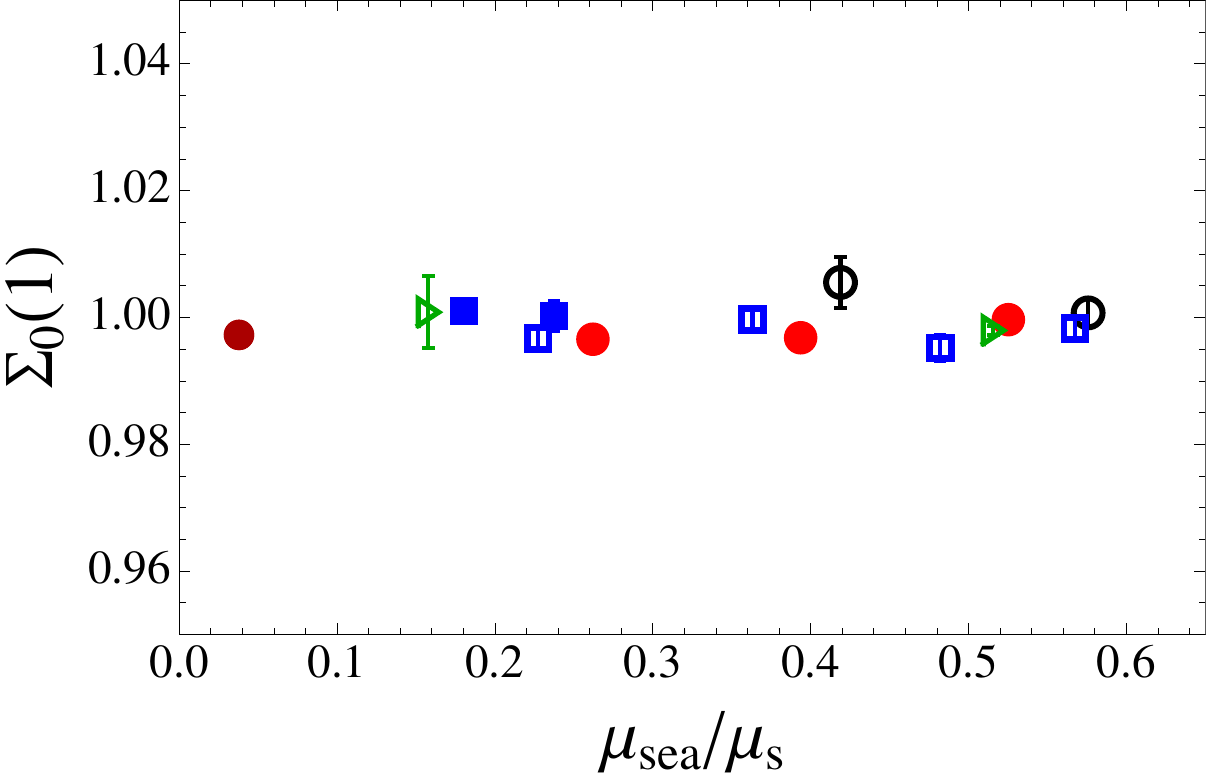}
\includegraphics[scale=.65]{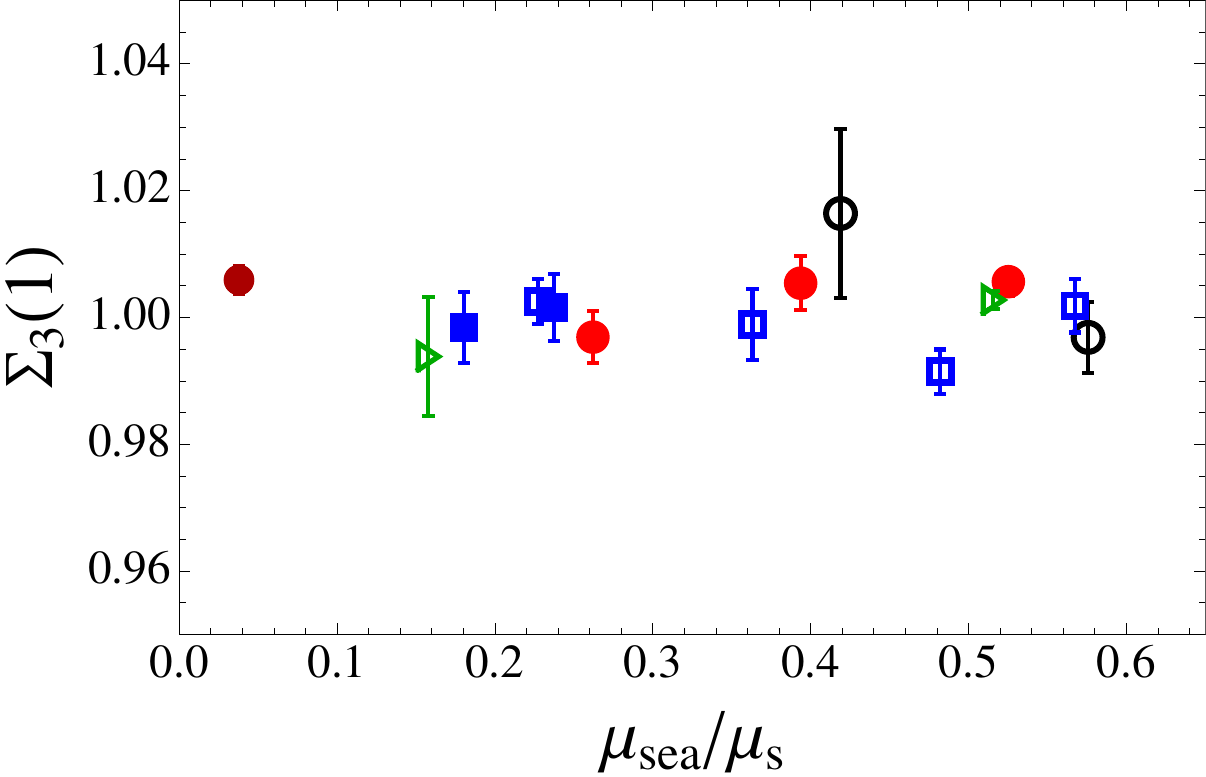}
\caption{\label{fig:01} \sl Values of $\Sigma_0$ and $\Sigma_3$ as obtained on all of the lattices used in this work is shown as a function of the light sea quark mass (divided by the physical strange quark mass). Different symbols are used to distinguish the lattice data obtained at different lattice spacings: { \Large{$\bm \circ$}} for $\beta=3.80$,  { \color{blue}$\bm\square$} for $\beta=3.90$ ($24^3$), ${\color{blue} \large \blacksquare}$  for $\beta=3.90$ ($32^3$),
 {\color{red} \LARGE \textbullet} for $\beta=4.05$,  and  ${\color{Green} \bm{ \rhd}}$ for $\beta=4.20$. The result of continuum extrapolation is also indicated at the point corresponding to the physical $\mu_{ud}/\mu_s\equiv m_{ud}/m_s = 0.037(1)$~\cite{charm1}.}
\end{figure}

We then computed the ratios of the form factors $\gone$ obtained at each two successive heavy valence quarks as indicated in eq.~(\ref{eq:bigSigma}). Importantly, a strong cancellation of statistical errors leads to very accurate $\Sigma_k$'s. 
The values of all $\Sigma_k$'s are presented in tab.~\ref{tab:0x}. In fig.~\ref{fig:01} we illustrate the situation for two values of $k$. From these plots we can see that our lattice data exhibit very little or no dependence on the light sea quark mass, nor on the lattice spacing. Note also that for larger heavy quark masses the errors on $\Sigma_k$ are larger, and therefore the corresponding continuum value $\sigma_k$ will have larger error as well. 
\begin{table*}[h!!]
\begin{ruledtabular}
\begin{tabular}{|c|ccccccccc|}   
{\phantom{\huge{l}}}\raisebox{-.2cm}{\phantom{\Huge{j}}}
Ensemble & $\Sigma_0(1)$ & $\Sigma_1(1)$& $\Sigma_2(1)$& $\Sigma_3(1)$& $\Sigma_4(1)$& $\Sigma_5(1)$& $\Sigma_6(1)$& $\Sigma_7(1)$& $\Sigma_8(1)$  \\ \hline
{\phantom{\huge{l}}}\raisebox{-.2cm}{\phantom{\Huge{j}}}
I      &   1.001(3) &   1.001(4) &  1.001(4) &   0.997(5) & 0.993(9) &  0.981(17) & 0.934(35) & 0.762(81) & 0.020(334)  \\
{\phantom{\huge{l}}}\raisebox{-.2cm}{\phantom{\Huge{j}}}
II      &   1.005(4) &   1.010(7) &  1.011(8) &   1.012(12) & 1.007(22) &  1.024(39) & 1.011(86) & 0.859(242) & 0.121(717)  \\
{\phantom{\huge{l}}}\raisebox{-.2cm}{\phantom{\Huge{j}}}
III      &   0.997(2) &   1.004(4) &  1.002(3) &   0.997(6) & 1.000(6) &  0.998(10) & 1.003(19) & 1.028(35) & 1.121(90)  \\
{\phantom{\huge{l}}}\raisebox{-.2cm}{\phantom{\Huge{j}}}
IV    &   0.995(2) &   0.994(4) &  0.994(3) &   0.991(4) & 0.989(4) &  0.986(8) & 0.970(17) & 0.960(40) & 0.892(126)  \\
{\phantom{\huge{l}}}\raisebox{-.2cm}{\phantom{\Huge{j}}}
V     &   0.999(2) &   1.000(3) &  1.000(5) &   0.999(6) & 0.999(6) &  1.002(8) & 1.002(16) & 1.031(47) & 1.115(140)  \\
{\phantom{\huge{l}}}\raisebox{-.2cm}{\phantom{\Huge{j}}}
VI     &   0.996(2) &   1.003(4) &  1.002(3) &   1.003(3) & 1.000(6) &  0.996(11) & 0.988(25) & 1.004(64) & 1.102(194)  \\
{\phantom{\huge{l}}}\raisebox{-.2cm}{\phantom{\Huge{j}}}
VII     &   1.000(3) &   1.001(3) &  1.002(4) &   1.002(5) & 1.002(7) &  1.006(9) & 1.021(14) & 1.039(36) & 1.037(133)  \\
{\phantom{\huge{l}}}\raisebox{-.2cm}{\phantom{\Huge{j}}}
VIII     &   1.002(2) &   1.002(4) &  1.003(5) &   1.000(5) & 1.020(20) &  0.974(23) & 0.976(16) & 0.932(33) & 0.851(85)  \\
{\phantom{\huge{l}}}\raisebox{-.2cm}{\phantom{\Huge{j}}}
IX     &   0.996(2) &   1.004(4) &  0.998(4) &   0.998(5) & 0.995(5) &  0.989(8) & 0.979(13) & 0.970(24) & 0.926(51)  \\ 
{\phantom{\huge{l}}}\raisebox{-.2cm}{\phantom{\Huge{j}}}
X     &   0.997(2) &   0.996(3) &  1.005(4) &   0.999(4) & 1.005(6) &  1.007(10) & 1.013(16) & 1.027(30) & 1.069(76)  \\ 
{\phantom{\huge{l}}}\raisebox{-.2cm}{\phantom{\Huge{j}}}
XI     &   0.998(1) &   1.009(2) &  1.004(2) &   1.007(2) & 1.008(3) &  1.011(4) & 1.016(7) & 1.025(14) & 1.025(27)  \\ 
{\phantom{\huge{l}}}\raisebox{-.2cm}{\phantom{\Huge{j}}}
XII     &   0.996(1) &   1.010(3) &  1.000(2) &   1.007(2) & 1.005(1) &  1.005(3) & 1.008(3) & 1.008(3) & 1.018(7)  \\ 
{\phantom{\huge{l}}}\raisebox{-.2cm}{\phantom{\Huge{j}}}
XIII     &   1.001(6) &   1.000(9) &  0.997(8) &   0.994(11) & 0.988(12) &  0.981(15) & 0.970(21) & 0.954(34) & 0.925(93)  \\ 
\end{tabular}
{\caption{\footnotesize  \label{tab:0x} Results of the ratios of the form factor $\gone$ computed at successive heavy quark masses according to eq.~(\ref{eq:bigSigma}), as computed on all of our ensembles of gauge field configurations.
.}}
\end{ruledtabular}
\end{table*}

\begin{figure}[t!]
\centering
\hspace*{-1.6mm}\includegraphics[scale=.58]{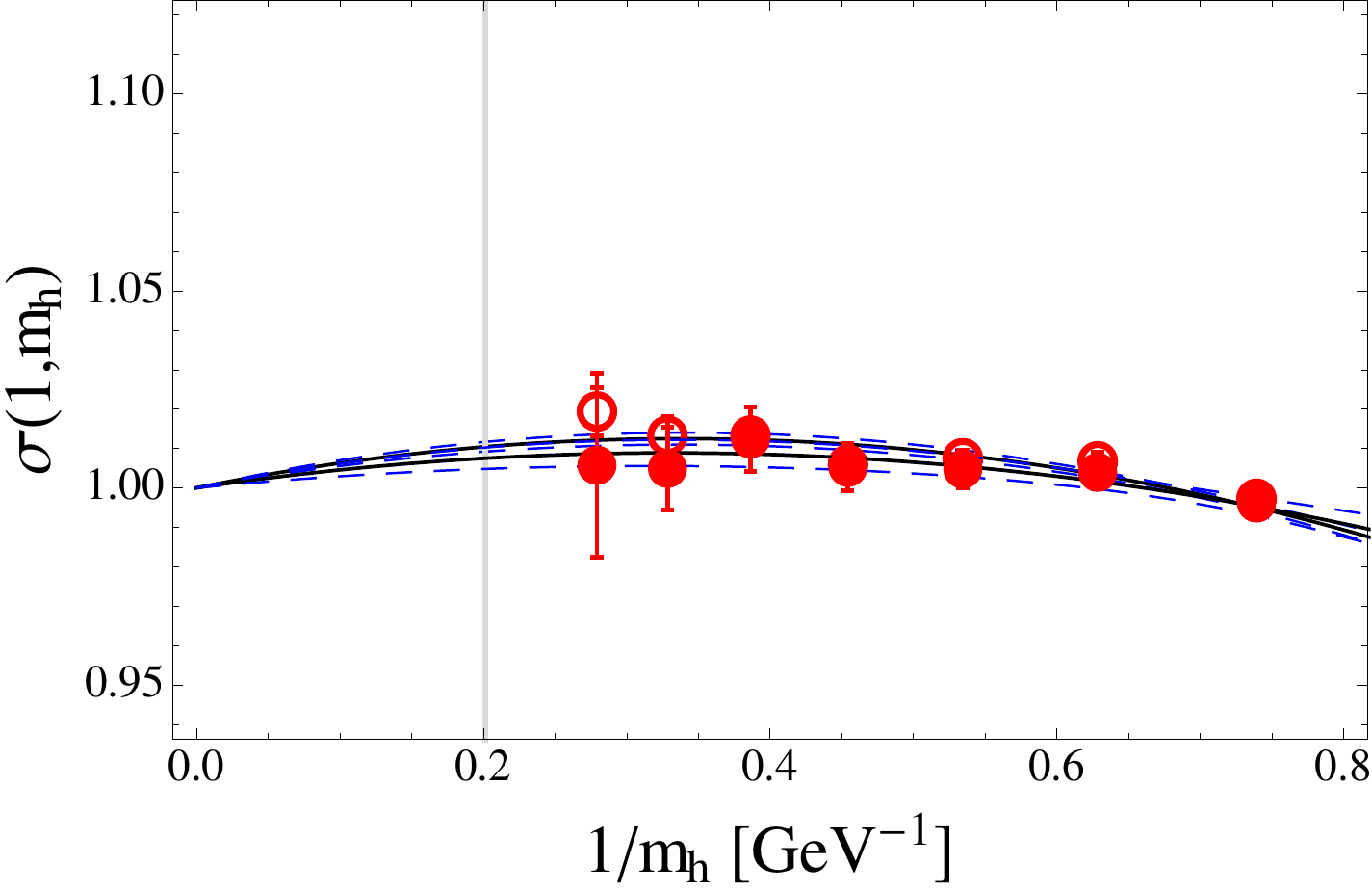}
\caption{\label{fig:02} \sl We show our data for $\sigma_k(1) = \sigma(1,m_h)$ with $m_h=\lambda^{k+1}m_c$, and show the result of the fit in $1/m_h$ to the form given in eq.~(\ref{fitSIGMA}) as a function of the inverse heavy quark mass with $m_h=\lambda^{k+1}m_c$. Filled symbols correspond to $\sigma(1,m_h)$ extrapolated to the continuum limit by using eq.~(\ref{form:1}) with all parameters free, whereas the empty symbols refer to the results obtained by imposing $\beta_k=0$. Fitting curves of the central values, together with the bounds (dashed lines), are also displayed. The gray vertical line indicates to point corresponding to the inverse of the physical $b$-quark mass.  }
\end{figure}

The extrapolation of $\Sigma_k(1)$ to the continuum limit is performed by using the following form
\bea\label{form:1}
\Sigma_k(1) = \alpha_k + \beta_k \frac{m_{\rm sea}}{m_s} + \gamma_k \l({a\over a_{\beta=3.9}}\r)^2,
\eea
and then identify
\bea
\sigma_k(1) =  \lim_{\substack{a\to 0 \\ m_{\rm sea}\to  m_{ud} }}   \Sigma_k(1)\,,
\eea
where $m_{ud}$ stands for the average of the physical up and down quark masses computed on the same lattices~\cite{charm1}. 
As anticipated from fig.~\ref{fig:01} the values of $\beta_k$ and $\gamma_k$, as obtained from the fit of our data to eq.~(\ref{form:1}), are consistent with zero. The resulting $\sigma_k(1)=\sigma(1, \lambda^{k+1}m_c)\equiv \sigma(1,m_h)$ are given in tab.~\ref{tab:02}. Since our data do not exhibit a dependence on the sea quark mass we also attempted extrapolating $\Sigma_k(1) \to \sigma_k(1)$ by imposing $\beta_k=0$ in eq.~(\ref{form:1}). Results for the first few $\sigma_k(1)$ remain practically indistinguishable from those obtained by letting $\beta_k$ as a free fit parameter. For higher masses, namely for $\Sigma_{4-8}(1)$, the results of two continuum extrapolations remain compatible but the error bars in the case of a free $\beta_k$ are considerably larger. The problem of larger errors for large quark masses is circumvented by the interpolation formula~(\ref{form:2}). Clearly the data with larger error bars become practically irrelevant in the fit because the intercept of the fit is fixed to unity by the heavy quark symmetry.  In other words, the result of interpolation to $\sigma(1,m_b)$ remains unchanged regardless of whether we include our results for $\sigma_{4-8}(1)$ in the fit or not. We stress again that instead of extrapolating ${\cal G}(1, m_h, m_c)$ in inverse heavy quark mass to the physically interesting point [${\cal G}(1, m_b, m_c)$] one interpolates $\sigma(1,m_h)$ to $\sigma(1,m_b)$ since $\ds{\lim_{m_h\to\infty}}\sigma(m_h)=1$. In practice we identify  $\sigma_k(1) = \sigma(1,\lambda^{k+1} m_c)$, and then like suggested in eq.~(\ref{form:2}), fit our results to  
\bea\label{fitSIGMA}
\sigma(1,m_h) = 1 +  \frac{s_1}{m_h} +  \frac{s_2}{m_h^2} \,,
\eea
which is illustrated in fig.~\ref{fig:02}. 
\begin{table*}
\newpage \begin{ruledtabular}
\begin{tabular}{|c|ccccccccc|} 
{\phantom{\huge{l}}} \raisebox{-.2cm} {\phantom{\huge{j}}}
$k$ &   0 & 1 & 2 & 3 & 4 & 5 & 6 & 7 & 8   \\ \hline
{\phantom{\huge{l}}} \raisebox{-.2cm} {\phantom{\huge{j}}}
$1/m_h\ [\gev ]$ &   0.739 & 0.629  & 0.534 & 0.454 & 0.386 & 0.328 & 0.279 & 0.237 & 0.202  \\  
{\phantom{\huge{l}}} \raisebox{-.2cm} {\phantom{\huge{j}}}          
$\sigma(1,m_h)_{\beta_k=0}$ &   0.991(1) & 1.007(2)  & 1.008(2) & 1.006(2) & 1.013(2) & 1.013(5) & 1.019(6) & 1.022(11) & 1.060(31)  \\  
{\phantom{\huge{l}}} \raisebox{-.2cm} {\phantom{\huge{j}}}          
$\sigma(1,m_h)_{\beta_k \ \rm free }$ &   0.996(2) & 1.004(4)  & 1.005(5) & 1.005(6) & 1.012(8) & 1.005(10) & 1.006(23) & 0.975(48) & 0.853(111)  \\  
\end{tabular}
{\caption{  \label{tab:02} \sl
Results of the continuum extrapolation of $\Sigma_k(1)$ to $\sigma_k(1)=\sigma(1,\lambda^{k+1}m_c)\equiv \sigma(1,m_h)$ using eq.~(\ref{form:1}). Results of extrapolation with $\beta_k$ as a free parameter are shown separately from those in which the observed independence on the sea quark mass is imposed in the fit~(\ref{form:1}) by setting $\beta_k=0$. 
}}
\end{ruledtabular}
\end{table*}
We then proceed as in eq.~(\ref{eq:main1}) and obtain 
\bea\label{eq:g1}
&&\gone = 1.073(17) \qquad (\beta_k=0)\,,\nn\\
&&\gone = 1.052(46) \qquad (\beta_k\neq 0)\,.
\eea
The first (more accurate) result agrees with the only existing unquenched lattice QCD result, obtained for the light non-strange spectator quark~\cite{Okamoto:2004xg}. To calculate our results in eq.~(\ref{eq:g1}) no renormalization constant was actually needed. This is convenient but not particularly beneficiary for our computation since the vector current renormalization constants have been computed non-perturbatively in ref.~\cite{Constantinou:2010gr} to a very good accuracy (cf. values listed in tab.~\ref{tab:01}). Therefore, we were able to perform several checks and instead of starting from the elastic form factor $ {\cal G}(1, m_c,m_c) = 1$, we could have started from a $k<n$ to compute $ {\cal G}(1, \lambda^{k+1} m_c,m_c)$ in the continuum limit, and then applied $\sigma_{k+1}\dots\sigma_n$ to reach the $b$-quark mass. For example, by using $k=3$, 
\begin{align}
{\cal G}(1,m_b,m_c)&=\sigma_8\sigma_7\sigma_6\sigma_5 \sigma_4  \ {\cal G}(1,\lambda^4 m_c,m_c) \nn\\
&= 1.059(47),
\end{align}
in the case with $\beta_k\neq 0$. This results is obviously completely consistent with the number given in eq.~(\ref{eq:g1}). To get the above result we also needed to perform a continuum extrapolation of ${\cal G}(1,\lambda^4 m_c,m_c)$ by using the expression analogous to the one shown in eq.~(\ref{form:1}). We checked and observed that our lattice data for the form factor are also independent on the light sea quark when the valence quark masses are fixed, a behavior very similar to what is shown in fig.~\ref{fig:01}. 
Furthermore we checked that, after adding the cubic term in $1/m_h$ to eq.~(\ref{fitSIGMA}), the resulting  $\gone = 1.047(61)$, remains fully consistent with our main result given in eq.~(\ref{eq:g1}). Although th finite volume effects are not expected to affect the quantities computed in this paper, they could appear when the dynamical (sea) quark mass is lowered. In order to check for that effect we can compare our results obtained on the ensembles VI and VII which differ by the volume. The situation shown in fig.~\ref{fig:01}  is a generic illustration of the situation we see with all the other quantities: the form factors are completely insensitive to a change of the lattice volume.   
All these checks suggest that our result (\ref{eq:g1}) obtained by using $\beta_k$ as a free parameter, remains stable and we take it for our final result, namely 
\bea\label{gone-2}
\boxed{\gone = 1.052(46) .}
\eea

Finally, we repeated the whole computation for the non-strange case, i.e. by keeping the sea and valence light quarks degenerate in mass. We obtained $\gone = 1.079(29)$ and $\gone = 1.033(95)$, corresponding to $\beta_k=0$ and $\beta_k\neq 0$ respectively. This latter number is not helpful in reducing the error bar of $\vert V_{cb}\vert$ extracted  from $B\to D\mu\nu$ decays. It shows, however, that the method employed in this work can be used to get a percent precision of $\gone$ even in the non-strange case provided the statistical quality of the data is substantially improved.  Note also that our $\gone$ in eq.~(\ref{gone-2}) agrees with the result obtained by the expansion around the BPS limit in ref.~\cite{bps}.

We end this discussion with a comment concerning the non-zero recoil situation ($w\neq 1$). The ana\-ly\-sis is essentially the same as in the zero-recoil case described above. From the correlation functions~(\ref{eq:three-tw}) and by using the projector ${\mathbb P}_\mu^+$~(\ref{proj}) we get the form factor $f_+(q^2)$ which is proportional to the desired ${\cal G}(w,\lambda^k m_c, m_c)$, cf. eqs.~(\ref{eq:f2h},\ref{eq:Hw2}).  The observations made in the analysis of $\gone$ concerning the independence on the light sea quark mass and on the lattice spacing remain true after switching from $w=1$ to $w\neq 1$. The values are given in tab.~\ref{tab:03}, where we again report our results both in the case in which the parameter $\beta_k$ in the continuum extrapolation~(\ref{form:1}) is left free and in the case  in which $\beta_k=0$ is imposed.  The net effect in the latter case is that the resulting error is considerably smaller. 
\begin{table*}[h!!]
\begin{ruledtabular}
\begin{tabular}{|c|cc|cc|cc|} 
{\phantom{\huge{l}}} \raisebox{-.2cm} {\phantom{\huge{j}}}
$w$~$\l(q^2_{B_s\to D_s} [\gev^2]\r)$ &  \multicolumn{2}{c|}{$\gw$} &  \multicolumn{2}{c|}{$\ds{f_0(q^2)\over f_+(q^2)}$} &  \multicolumn{2}{c|}{$\ds{f_T(q^2)\over f_+(q^2)}$}   \\ \hline 
                                                                     &   $\beta_k \neq 0$ &   $\beta_k = 0$ &   $\beta_k^\prime \neq 0$ &   $\beta_k^\prime = 0$&   $\beta_k^{\prime\prime} \neq 0$ &   $\beta_k^{\prime\prime}  = 0$\\  \hline
{\phantom{\huge{l}}} \raisebox{-.2cm} {\phantom{\huge{j}}}
1.  (11.54)&  1.052(47) & 1.073(17)  & - & - & - & -   \\  
{\phantom{\huge{l}}} \raisebox{-.2cm} {\phantom{\huge{j}}}
1.004 (11.46) &  1.052(47) & 1.075(16)  & 0.766(19) & 0.752(7) & 1.076(68) & 1.078(43)   \\  
{\phantom{\huge{l}}} \raisebox{-.2cm} {\phantom{\huge{j}}}
1.016 (11.20) &  1.029(49) & 1.063(15)  & 0.781(24) & 0.757(9) & 1.062(76) & 1.064(49)   \\  
{\phantom{\huge{l}}} \raisebox{-.2cm} {\phantom{\huge{j}}}
1.036 (10.79) &  1.044(51) & 1.034(17)  & 0.787(34) & 0.760(16) & 0.975(94) & 0.997(64)   \\  
{\phantom{\huge{l}}} \raisebox{-.2cm} {\phantom{\huge{j}}}
1.062 (10.23) &  0.986(57) & 1.004(20)  & 0.825(59) & 0.761(34) & 0.920(111) & 1.004(76)   \\  
\end{tabular}
{\caption{  \label{tab:03} \sl
All physically relevant results of this study: $\gw$ is the dominant form factor governing the hadronic matrix element relevant to $B_s\to D_s\ell \nu_\ell$ ($\ell \in\{e,\mu\}$) computed in the zero recoil region, $f_0(q^2)/f_+(q^2)$ is needed for $B_s\to D_s\tau \nu_\tau$ in the Standard Model and for all the leptons in the case of helicity enhanced contributions present in the models beyond Standard Model. The tensor form factor $f_T(q^2)$, also needed in some NP scenarios, is computed at $\mu = m_b$ in the $\msbar$ renormalization scheme.}}
\end{ruledtabular}
\end{table*}
Using the parameterization of ref.~\cite{CLN}, which takes into account the relation between the curvature and the slope of $\gw$, namely
\begin{align}
{\gw \over \gone} = 1 - 8 \rho^2 z &+ (51 \rho^2-10) z^2 \nn\\
 &- (252 \rho^2 - 84)z^3\,,
\end{align}
with $z=(\sqrt{w+1}-\sqrt{2})/(\sqrt{w+1}+\sqrt{2})$, one could attempt to extract the slope $\rho^2$ from our data. Knowing that the window of $w$'s we consider here is very short~(\ref{nonzero}), a clean determination of  $\rho^2$ would require very accurate values of $\gw$. In our case we only obtain $\rho^2=1.2(8)$, or in the case where we dismiss the dependence on the sea quark mass (when the errors on $\gw$ are smaller) we get $\rho^2=1.1(3)$, both being consistent with the experimentally established $\rho^2=1.19(4)(4)$~\cite{hfag1}. The same quality of result for $\rho^2$ is obtained if the data are fit to~\cite{orsay} 
\bea
{\gw \over \gone} = \l(\frac{2}{1+w} 
\r)^{2\rho^2}.
\eea

\section{Scalar and Tensor Form Factors\label{sec:bsm}}

Recently measured $B(B\to D\tau \nu_\tau)$ by the BaBar collaboration indicated about $2\sigma$-discrepancy with respect to the SM estimate, obtained by combining the measured $B(B\to D\mu \nu_\mu)$ and the known information about the scalar form factor~\cite{BABAR}. 
Since then, a number of studies appeared trying to explain that discrepancy by interpreting it as a potential signal of New Physics (NP)~\cite{BDpapers,BKT}. 
In the models with two Higgs doublets (2HDM), the charged Higgs boson can mediate the tree level processes, including $B\to D\ell \nu$, and considerably enhance the coefficient multiplying the scalar form factor in the decay amplitude. For that reason it becomes important to get a lattice QCD estimate of $f_0(q^2)$. 
Furthermore, the model independent considerations of NP also allow a possibility of having a non-zero tensor coupling, in which case one more form factor appears in the decay amplitude.  
The tensor form factor $f_T(q^2)$ is defined via,  
\begin{align}
\label{eq:fTT}
\langle D_s(k) \vert \bar b \sigma_{\mu\nu} &c \vert  B_s(p)\rangle  =  \nn\\
&- i
\left( p_\mu k_\nu -  k_\mu p_\nu \right) {2 \ f_T(q^2,\mu) \over m_{B_s}+m_{D_s}}, 
\end{align}
where the renormalization scale dependence reflects the fact that the tensor density in QCD is a logarithmically divergent operator. In what follows the $\mu$-dependence will be tacitly assumed. 
In this paper we report the result of the first lattice QCD computation of the tensor form factor in the region close to $q^2_{\rm max}$.

More specifically, in this section we compute 
\bea
R_0(q^2) = {f_0(q^2)\over f_+(q^2)}\,,\quad R_T(q^2) = {f_T(q^2)\over f_+(q^2)}\,,
\eea
which directly enter the expression for differential decay rate that can be found in eg. ref.~\cite{BKT}.  Since the form factors $f_{+,T}(q^2)$ are not accessible at $q^2_{\rm max}$ (zero recoil, $w=1$), we computed $R_{0,T}(w(q^2))$ at $w\gtrsim 1$.

\subsection{$R_0(q^2)$}
The extraction of $R_0(q^2)$ is practically straightforward. After applying the projectors ${\mathbb P}_\mu^{+,0}$~(\ref{proj}) to the matrix element extracted from the correlation functions~(\ref{eq:three-tw}), we combine them in the ratios $R_0(w, m_h, m_c) =R_0(w, \lambda^k m_c, m_c)$. Our goal is again to use the ratios of $R_0(w)$ computed at successive heavy quark masses, and then reach the point corresponding to the physically relevant $R_0(w, m_b, m_c)$ through interpolation in inverse heavy quark mass. To this end, we first form
\bea
\Sigma^0_k(w) ={  R_0(w, \lambda^{k+1} m_c, m_c,a^2) \over   R_0(w, \lambda^{k} m_c, m_c, a^2) }\,,
\eea
where we indicate the momentum transfer $w$, the masses of quarks entering the weak vertex ($m_c$, $m_h$), and the fact that $\Sigma_k^0(w)$'s are obtained at fixed lattice spacing, $a$.  In tab.~\ref{tab:0y} we present the results for $\Sigma^0_k(w)$ for a specific value of $w=1.016$. Before discussing the heavy quark mass dependence we need to extrapolate to the continuum limit, $\ds{\lim_{a\to 0}}\Sigma^0_k(w) = \sigma^0_k(w)$ by using a form similar to~(\ref{form:1}), 
\begin{align}\label{eq:cont0}
\Sigma_k^0(w,m_{\rm sea}, a^2) =& \alpha_k^\prime(w) + \beta_k^\prime(w) \frac{m_{\rm sea}}{m_s} \nn\\
&+ \gamma_k^\prime(w) \l({a\over a_{\beta=3.9}}\r)^2,
\end{align}
thus assuming the linear dependence on the dynamical (``sea") quark mass and on the square of the lattice spacing. Since we work with maximally twisted QCD on the lattice, the leading discretization errors are proportional to $a^2$~\cite{fr}. After inspection, we found again that the dependence of the form factors on the sea quark mass is indiscernible from our data and that our results depend very mildly on the lattice spacing. Since the dependence on the sea quark mass is negligible, we again consider the  continuum extrapolation by setting $\beta_k^\prime(w)=0$, separately from the case in which $\beta_k^\prime(w)$ are left as free parameters. The net effect is that the error on $\sigma^0_k(w)=\Sigma_k(w,0, 0)$ is considerably smaller in the case with $\beta_k^\prime(w)=0$ and the data better respect the heavy quark mass dependence. 
\begin{figure}[t!]
\centering
\hspace*{-4.6mm}\includegraphics[scale=.56]{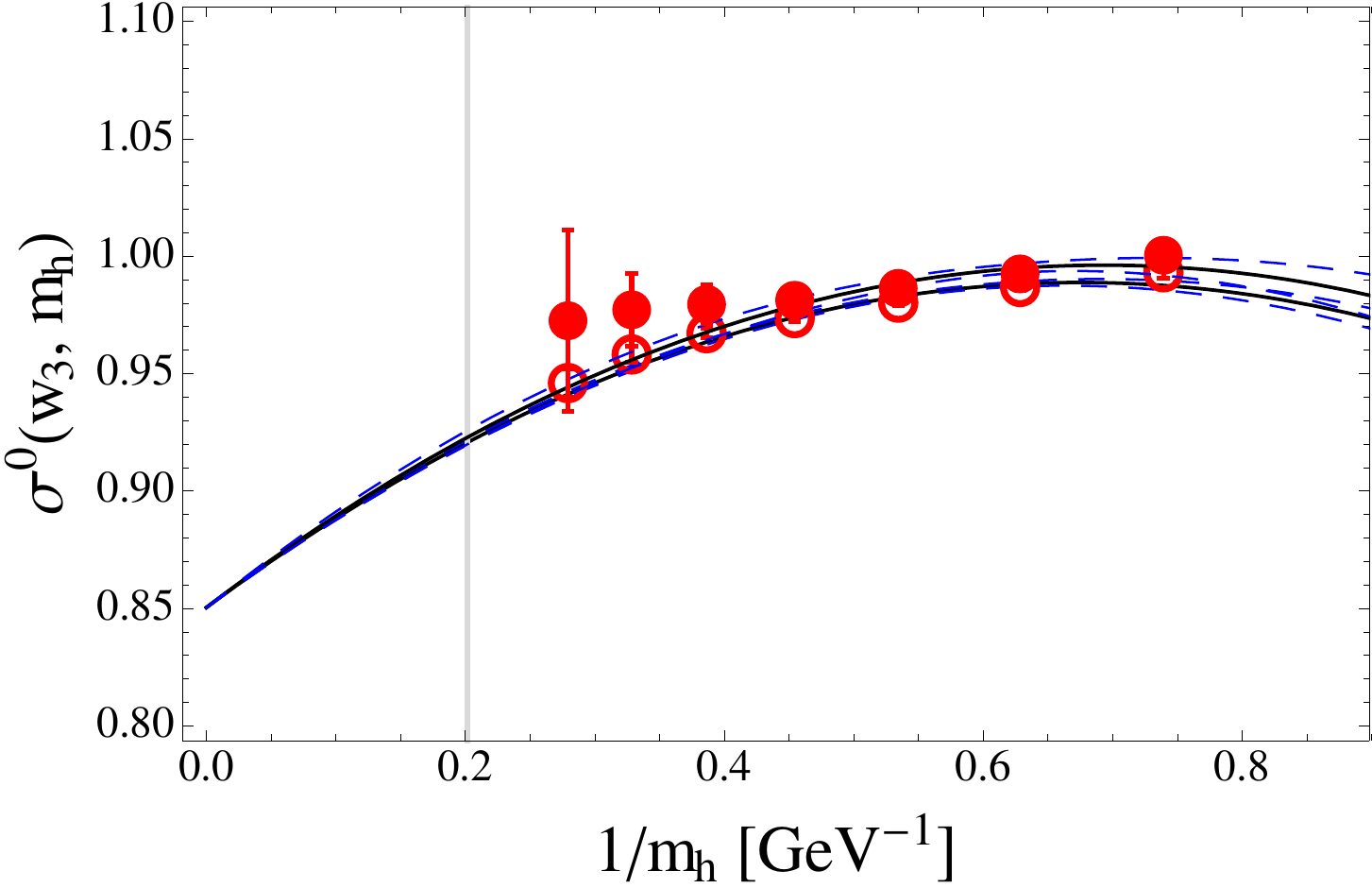}
\caption{\label{fig:04} \sl Fit of our data to eq.~(\ref{fit:0plus}). Empty symbols denote the results computed in the continuum by setting $\beta_k^\prime(w)=0$ in eq.~(\ref{eq:cont0}). Filled symbols correspond to the results obtained after allowing $\beta_k^\prime(w)\neq 0$ in eq.~(\ref{eq:cont0}). Plotted are the data with $w=w_3=1.016$ }
\end{figure}
\begin{table*}
\begin{ruledtabular}
\begin{tabular}{|c|cccccccc|}   
{\phantom{\huge{l}}}\raisebox{-.2cm}{\phantom{\Huge{j}}}
Ensemble &  $\Sigma^0_1(w_3)$& $\Sigma_2(w_3)$& $\Sigma^0_3(w_3)$& $\Sigma^0_4(w_3)$& $\Sigma^0_5(w_3)$& $\Sigma^0_6(w_3)$& $\Sigma^0_7(w_3)$& $\Sigma^0_8(w_3)$  \\ \hline
{\phantom{\huge{l}}}\raisebox{-.2cm}{\phantom{\Huge{j}}}
I      &   0.986(3) &   0.986(2) &  0.984(3) &   0.982(6) & 0.983(11) &  0.992(27) & 1.054(97) & 0.9(1.9)   \\
{\phantom{\huge{l}}}\raisebox{-.2cm}{\phantom{\Huge{j}}}
II      &   0.993(6) &   0.991(4) &  0.988(5) &   0.987(8) & 0.990(18) &  1.015(48) & 1.181(23) & -0.5(3.2)   \\
{\phantom{\huge{l}}}\raisebox{-.2cm}{\phantom{\Huge{j}}}
III      &  0.988(4) &   0.986(3) &  0.981(4) &   0.975(5) & 0.969(7) &  0.965(13) & 0.964(32) & 0.974(90)   \\
{\phantom{\huge{l}}}\raisebox{-.2cm}{\phantom{\Huge{j}}}
IV    &   0.996(4) &   0.992(2) &  0.987(2) &   0.982(4) & 0.977(7) &  0.974(16) & 0.991(35) & 1.064(94)  \\
{\phantom{\huge{l}}}\raisebox{-.2cm}{\phantom{\Huge{j}}}
V     &   0.991(4) &   0.988(3) &  0.983(3) &   0.976(4) & 0.966(7) &  0.947(13) & 0.904(27) & 0.810(64)  \\
{\phantom{\huge{l}}}\raisebox{-.2cm}{\phantom{\Huge{j}}}
VI     &   0.991(5) &   0.989(3) &  0.986(3) &   0.983(3) & 0.981(5) &  0.978(11) & 0.974(27) & 0.957(70)   \\
{\phantom{\huge{l}}}\raisebox{-.2cm}{\phantom{\Huge{j}}}
VII     &    1.000(4) &   0.994(2) &  0.989(2) &   0.984(3) & 0.980(4) &  0.980(9) & 0.994(19) & 1.048(48)   \\
{\phantom{\huge{l}}}\raisebox{-.2cm}{\phantom{\Huge{j}}}
VIII     &   0.995(4) &   0.991(2) &  0.988(2) &   0.984(3) & 0.983(5) &  0.988(11) & 1.020(30) & 1.169(128)  \\
{\phantom{\huge{l}}}\raisebox{-.2cm}{\phantom{\Huge{j}}}
IX     &   0.992(3) &   0.989(2) &  0.985(3) &   0.981(4) & 0.978(5) &  0.975(9) & 0.967(20) & 0.946(55)   \\ 
{\phantom{\huge{l}}}\raisebox{-.2cm}{\phantom{\Huge{j}}}
X     &   0.999(7) &   0.9892(5) &  0.987(5) &   0.983(7) & 0.980(11) & 0.980(20) & 0.984(43) & 0.99(12)   \\ 
{\phantom{\huge{l}}}\raisebox{-.2cm}{\phantom{\Huge{j}}}
XI     &  0.991(2) &   0.988(1) &  0.983(1) &   0.978(2) & 0.972(2) &  0.967(4) & 0.966(8) & 0.981(22)   \\ 
{\phantom{\huge{l}}}\raisebox{-.2cm}{\phantom{\Huge{j}}}
XII     &   0.992(2) &   0.988(3) &  0.982(1) &   0.977(1) & 0.971(1) &  0.965(1) & 0.959(2) & 0.951(3)   \\ 
{\phantom{\huge{l}}}\raisebox{-.2cm}{\phantom{\Huge{j}}}
XIII     &    0.999(7) &   0.992(5) &  0.987(5) &   0.983(5) & 0.980(11) &  0.980(20) & 0.984(43) & 0.99(12)  \\ 
\end{tabular}
{\caption{\footnotesize  \label{tab:0y} Results of the ratios of $R_0(w_3)$ computed at successive heavy quark masses according to eq.~(\ref{eq:cont0}), on all of our ensembles of gauge field configurations. $w_3=1.016$.
.}}
\end{ruledtabular}
\end{table*}

With several $\sigma^0_k(w)$ in hands, we need to discuss the heavy quark interpolation. 
We first discuss its value in the infinitely heavy quark mass limit. Using the heavy quark effective theory mass formula $m_{B_s,D_s}= m_{b,c} + \overline \Lambda + (\lambda_1+3\lambda_2)/m_{b,c}$~\cite{manohar}, 
we can consider  the ratio of form factors given in eq.~(\ref{eq:f2h}), knowing that $h_+(w)$ scales as a constant with inverse heavy quark mass. One then deduces that, for the charm quark 
fixed to its physical value, 
\bea
R_0(w, m_h,m_c) \propto 1/m_h\,.
\eea
Equivalently, $\widetilde R_0(w, m_h,m_c) = m_h R_0(w, m_h,m_c)$, scales as a constant in the heavy quark mass limit, and the 
the corresponding $\widetilde \sigma_0(w)$ is then be described by a form similar to eq.~(\ref{form:2}) and the physically relevant $\widetilde R_0(w, m_b)$ could be obtained from 
\begin{align}
{\phantom{\huge{l}}} \raisebox{-.4cm} {\phantom{\huge{j}}}
\widetilde R_0(w, m_b)& = \widetilde \sigma^0_n \dots \widetilde\sigma^0_{k+1} \widetilde \sigma^0_k\ \widetilde  R_0(w, \lambda^k m_c) \ .
\end{align}
We can also rewrite the above formula in terms of $R_0(w)$, as
\begin{align}
{\phantom{\huge{l}}} \raisebox{-.4cm} {\phantom{\huge{j}}}
&\lambda^{n+1}  m_c R_0(w, m_b) = \widetilde \sigma^0_n \dots \widetilde\sigma^0_{k+1} \widetilde \sigma^0_k \lambda^{k} m_c R_0(w, \lambda^k m_c)\nn\\
& R_0(w, m_b)  =\underbrace{\lambda^{-n+k-1} \widetilde \sigma^0_n \dots \widetilde\sigma^0_{k+1} \widetilde \sigma^0_k}_{\ds \sigma^0_n \dots \sigma^0_{k+1} \sigma^0_k} R_0(w, \lambda^k m_c),
\end{align}
and therefore $ \sigma^0_k = \widetilde \sigma^0_n/\lambda$. In other words the interpolation formula to be used in this case is, 
\bea\label{fit:0plus}
\sigma^0(w,m_h) =  \frac{1}{\lambda} +  \frac{s_1(w)}{m_h} +  \frac{s_2(w)}{m_h^2}\,, 
\eea
where, again, our $\lambda=1.176$. An illustration of that interpolation is provided in fig.~\ref{fig:04} for one specific case of $w$. 
We see that our results obtained by assuming the independence of $R_0(w)$ on the sea quark scale better with the heavy quark mass than those  
obtained by letting the parameter $\beta_k^\prime(w)$ in eq.~(\ref{eq:cont0}) free, although the two are compatible within the error bars. 

\begin{figure}[t!]
\centering
\hspace*{-4.6mm}\includegraphics[scale=.56]{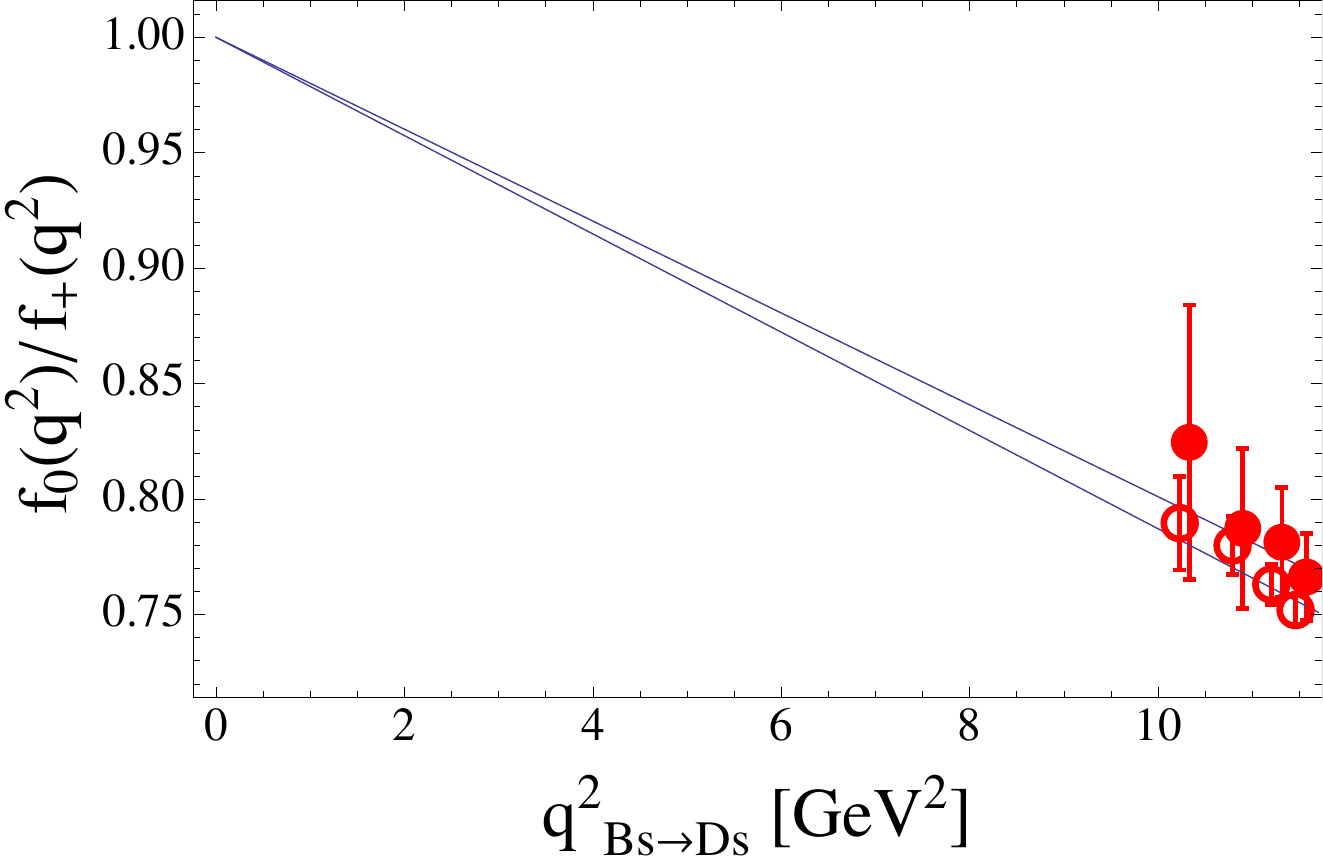}
\caption{\label{fig:05}\sl  Results for $R_0(q^2)=f_0(q^2)/f_+(q^2)$ presented in this paper in the case of $B_s\to D_s$ are linearly fit to the form $R_0(q^2)=1-\alpha q^2$. 
As in the previous plots, the empty/filled symbols correspond to the results obtained with $\beta_k^\prime(w)= 0$/ $\beta_k^\prime(w \neq 0$ in eq.~(\ref{eq:cont0}). }
\end{figure}

Our final results for $f_0(q^2)/f_+(q^2)$ with $q^2= q^2_{\rm max} - 2 m_{B_s} m_{D_s}(w-1)$, are given in tab.~\ref{tab:03}. Knowing that the form factors satisfy the constraint $f_0(0)=f_+(0)$, one can then attempt to fit linearly in $q^2$, 
as $R_0(q^2) = 1 - \alpha q^2$. From the results obtained with $\beta_k=0$ we obtain $\alpha=0.021(1)~\gev^{-2}$, while from the data with $\beta_k$ free, we get  $\alpha=0.020(1)~\gev^{-2}$. This is illustrated in fig.~\ref{fig:05}. 

It is interesting to note that these results are consistent with the values that can be obtained from the results quoted in recent literature (in the non-strange case). More specifically, from the lattice results of ref.~\cite{nazarioA} one finds $\alpha=0.020(1)\ \gev^{-2}$,  while from those reported in ref.~\cite{Bailey:2012rr} one finds $\alpha=0.022(1)\ \gev^{-2}$. Recent QCD sum rule analyses give  $\alpha=0.021(2)\ \gev^{-2}$~\cite{qsr}. 

Note also that near zero recoil, the central value of our result $R_0(q^2)=0.77(2)$, coincides with the quark model results of refs.~\cite{galkin, Melikhov:2000yu}.

\subsection{$R_T(q^2)$}

To our knowledge, there is no QCD based determination of the $B_{(s)}\to D_{(s)}$ transition tensor form factor. The only existing result is the one presented in ref.~\cite{Melikhov:2000yu} for the nonstrange case ($B_{ud}\to D_{ud}\ell \nu_\ell$) in which a constituent quark model has been employed. That obviously did not allow to keep track of the QCD anomalous dimension. 
However, as we shall see, their result [$f_T(q^2)/f_+(q^2)=1.03(1)$] is rather close to what we obtain from our lattice simulations. 
Furthermore, in ref.~\cite{Melikhov:2000yu} it was found that this ratio  $R_T(q^2)=f_T(q^2)/f_+(q^2)$ is a flat function of $q^2$.

\begin{figure}[t!]
\centering
\hspace*{-4.6mm}\includegraphics[scale=.56]{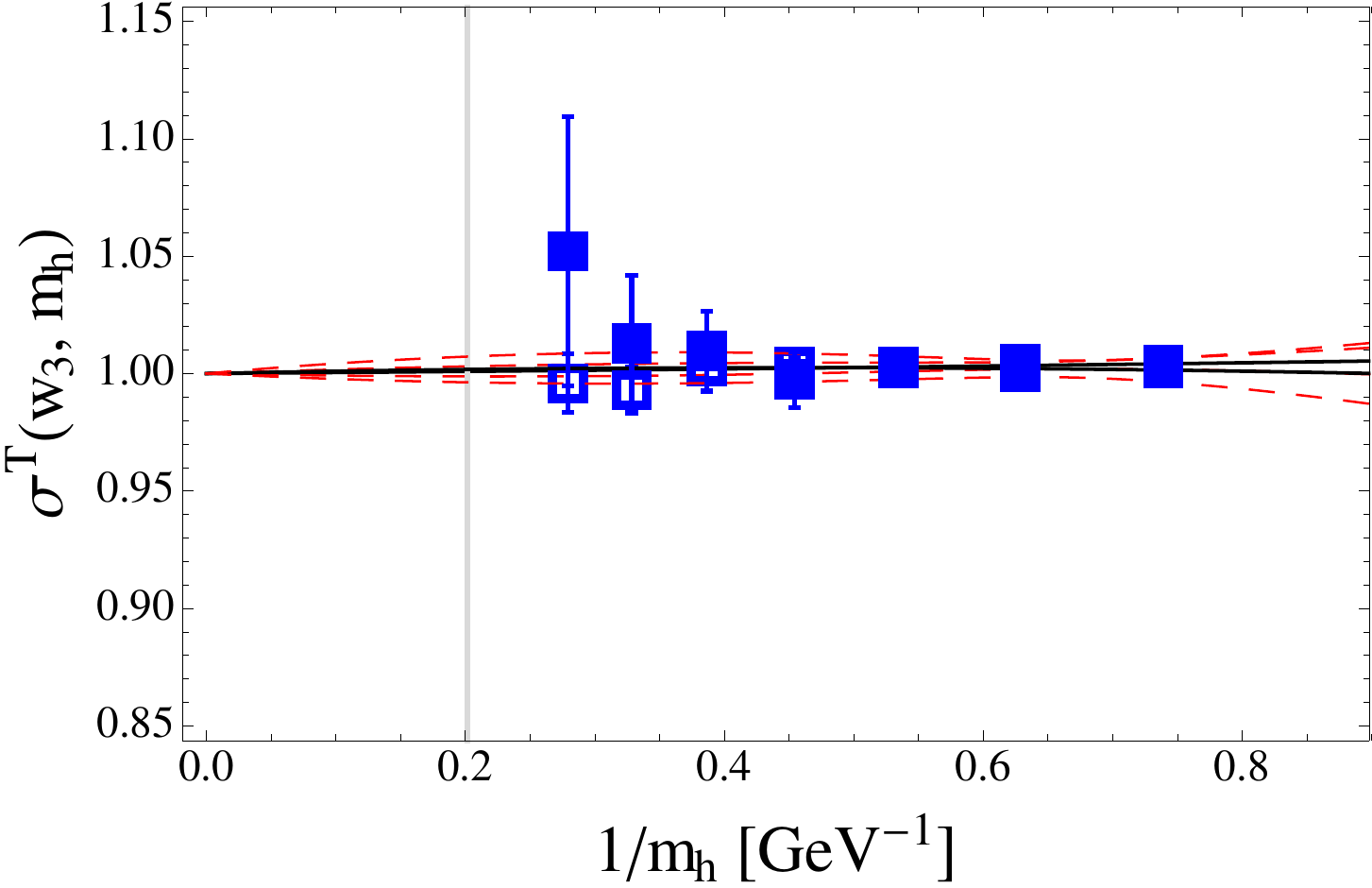}
\caption{\label{fig:06} \sl Plot analogous to fig.~\ref{fig:04} but for the case of  $\sigma^T(w,m_h)$ as given in eq.~(\ref{fit:0plus}). The vertical gray line corresponds to the inverse $b$-quark mass. }
\end{figure}

On the lattice, the extraction of the form factor $f_T(q^2)$ is completely analogous to what we explained in the previous sections for $f_+(q^2)$ and $f_0(q^2)$. Heavy quark behavior of $f_T(q^2)$ is similar to that of $f_+(q^2)$, 
which is simple to see after applying the heavy quark equation of motion to the $b$-quark,~\footnote{We stress again that the $c$-quark mass in our simulations is always kept fixed to its physical value.}  $\frac{1+\slash{\!\!\! v}}{2} b = b$, which in the $b$ rest frame reads $\gamma_0 b=b$, and therefore 
$\bar c \sigma_{0i}b = -i \bar c \gamma_i b$, so that the heavy quark behavior of the form factor $f_T(q^2)$ resembles that of $f_+(q^2)$.  
We again define the ratios computed at two successive quark masses that differ by a factor of $\lambda$, 
\bea
\Sigma^T_k(w) ={  R_T(w, \lambda^{k+1} m_c, m_c,a^2) \over   R_T(w, \lambda^{k} m_c, m_c, a^2) }\,,
\eea
which we then extrapolate to the continuum limit by using 
\begin{align}\label{eq:contT}
\Sigma_k^T(w,m_{\rm sea}, a^2) =& \alpha_k^{\prime\prime}(w) + \beta_k^{\prime\prime}(w) \frac{m_{\rm sea}}{m_s} \nn\\
&+ \gamma_k^{\prime\prime}(w) \l({a\over a_{\beta=3.9}}\r)^2.
\end{align}
Like in the previous cases, we observe that $\Sigma^T_k(w)$ does not depend on the sea quark mass and its dependence on lattice spacing is insignificant within our error bars. 
For that reason we made the continuum extrapolation by imposing $\beta_k^{\prime\prime}(w)=0$ and by leaving $\beta_k^{\prime\prime}(w)$ as a free parameter. Results of that extrapolation, $\sigma^T_k(w)\equiv  \sigma^T (w, m_h, m_c)$ with $m_h=\lambda^{k+1}m_c$, are then interpolated in heavy quark mass to the $b$-quark, according to,
\bea\label{fit:Tplus}
\sigma^T(w,m_h) =  1 +  \frac{\tilde s_1(w)}{m_h} +  \frac{\tilde s_2(w)}{m_h^2}\,, 
\eea
which is shown in fig.~\ref{fig:06}.
As in the case of $R_0(q^2)$, we also here need to extrapolate one of the ratios $R_T(q^2)$ to the continuum limit. We checked that for either $k=2$, or $3$ or $4$ we end up with completely consistent results for
\bea
R_T(w, m_b) = \sigma^T_k(w)\dots \sigma_8^T(w) \ R_T(w, \lambda^{k+1}m_c).
\eea
The results given in tab.~\ref{tab:03} are obtained by choosing $k=3$. Furthermore, we included the evolution of the tensor density from $\mu=2\ \gev$, at which the renormalization constants have been computed, to  
$\mu =m_b$ by using 
\bea
f_T(q^2,\mu) = \frac{c_T(\mu)}{c_T(\mu_0)} f_T(q^2,\mu_0)\,
\eea
where, 
\begin{align}
&c_T^{\rm lo}(\mu) = a_s(\mu)^{\gamma_0/\beta_0}\,,\nn\\
&c_T^{\rm nlo}(\mu) = c_T^{\rm lo}(\mu)\l( 
1 + \frac{\gamma_1\beta_0-\gamma_0\beta_1}{\beta_0^2}  a_s(\mu)\r)\,,\nn\\
&c_T^{\rm nnlo}(\mu) = c_T^{\rm nlo}(\mu)  +\frac{1}{2} c_T^{\rm lo}(\mu)\l[ 
\l(\frac{\gamma_1\beta_0-\gamma_0\beta_1}{\beta_0^2}\r)^2 +\frac{ \gamma_2}{\beta_0}\r.\nn\\
&\qquad\qquad\qquad \l.  +\frac{ \beta_1^2\gamma_0}{\beta_0^3} - \frac{\beta_1\gamma_1 + \beta_2 \gamma_0}{\beta_0^2}
\r] a_s(\mu)^2\,,
\end{align}
and where we used $a_s(\mu)\equiv \alpha_s(\mu)/\pi$, for shortness. The anomalous dimension coefficients are known to three loops in perturbation theory  and for $\nf=4$ in  the $\msbar$ scheme their values are~\cite{Gracey}:
\bea
\gamma_0 = \frac{1}{3}, \; \gamma_1 =\frac{ 439}{216}, \; \gamma_2 =4.002,\nn\\
\beta_0= \frac{25}{12}, \; \beta_1 =\frac{ 77}{24}, \; \beta_2 =\frac{21943}{3456},\nn
\eea
where we also gave the first few $\beta$-function coefficients. Finally, in the computation we used $\Lambda^\msbar_{\nf=4}=296(10)$~MeV~\cite{bethke}.

As it can be seen from tab.~\ref{tab:03} the error on $f_T(q^2)/f_+(q^2)$ is getting larger for larger values of $w$. We are therefore unable to check on the flatness of  $R_T(q^2)$ that is valid in the infinitely heavy quark mass limit.  

\begin{table*}
\begin{ruledtabular}
\begin{tabular}{|c|cccccccc|}   
{\phantom{\huge{l}}}\raisebox{-.2cm}{\phantom{\Huge{j}}}
Ensemble & $\Sigma^T_1(w_3)$& $\Sigma_2^T(w_3)$& $\Sigma^T_3(w_3)$& $\Sigma^T_4(w_3)$& $\Sigma^T_5(w_3)$& $\Sigma^T_6(w_3)$& $\Sigma^T_7(w_3)$& $\Sigma^T_8(w_3)$  \\ \hline
{\phantom{\huge{l}}}\raisebox{-.2cm}{\phantom{\Huge{j}}}
I      &   0.992(6) &   0.988(8) &  0.985(11) &   0.983(18) & 0.988(31) &  1.024(61) & 1.21(18) & 0.3(4.6)   \\
{\phantom{\huge{l}}}\raisebox{-.2cm}{\phantom{\Huge{j}}}
II      &   0.986(11) &   0.979(14) &  0.966(20) &   0.942(31) & 0.899(57) &  0.82(13) & 0.5(5) & -1.8(9.7)   \\
{\phantom{\huge{l}}}\raisebox{-.2cm}{\phantom{\Huge{j}}}
III      &  1.006(4) &   1.006(5) &  1.005(7) &  1.003(11) & 1.000(18) &  0.994(32) & 0.984(53) & 0.942(95)   \\
{\phantom{\huge{l}}}\raisebox{-.2cm}{\phantom{\Huge{j}}}
IV    &   1.010(5) &   1.010(6) &  1.010(7) &   1.011(8) & 1.012(11) &  1.025(22) & 1.049(52) & 1.12(20)  \\
{\phantom{\huge{l}}}\raisebox{-.2cm}{\phantom{\Huge{j}}}
V     &   0.998(4) &   0.997(5) &  0.996(6) &   0.995(8) & 0.994(12) &  0.995(20) & 0.978(43) & 0.938(94)  \\
{\phantom{\huge{l}}}\raisebox{-.2cm}{\phantom{\Huge{j}}}
VI     &   0.996(4) &   0.993(5) &  0.989(7) &   0.985(10) & 0.980(17) &  0.972(28) & 0.944(50) & 0.86(11)   \\
{\phantom{\huge{l}}}\raisebox{-.2cm}{\phantom{\Huge{j}}}
VII     &    1.002(3) &   1.001(4) &  1.001(5) &   1.001(7) & 1.000(12) &  0.996(24) & 0.998(50) & 1.02(13)   \\
{\phantom{\huge{l}}}\raisebox{-.2cm}{\phantom{\Huge{j}}}
VIII     &   1.002(3) &   0.999(5) &  1.003(6) &   0.992(10) & 1.019(17) &  1.030(21) & 1.101(50) & 1.37(21)  \\
{\phantom{\huge{l}}}\raisebox{-.2cm}{\phantom{\Huge{j}}}
IX     &   0.998(5) &   0.997(6) &  0.996(7) &   0.996(10) & 0.999(13) &  1.007(18) & 1.023(32) & 1.080(69)   \\ 
{\phantom{\huge{l}}}\raisebox{-.2cm}{\phantom{\Huge{j}}}
X     &   0.988(12) &   0.982(14) &  0.972(19) &   0.958(25) & 0.936(36) & 0.896(58) & 0.817(159) & 0.62(50)   \\ 
{\phantom{\huge{l}}}\raisebox{-.2cm}{\phantom{\Huge{j}}}
XI     &  1.000(1) &   0.999(2) &  0.998(2) &   0.996(3) & 0.994(5) &  0.990(7) & 0.986(12) & 0.993(24)   \\ 
{\phantom{\huge{l}}}\raisebox{-.2cm}{\phantom{\Huge{j}}}
XII     &   1.003(1) &   1.002(1) &  1.001(2) &   1.001(2) & 1.001(3) &  0.998(5) & 0.998(5) & 0.993(8)   \\ 
{\phantom{\huge{l}}}\raisebox{-.2cm}{\phantom{\Huge{j}}}
XIII     &    0.988(11) &   0.982(15) &  0.972(19) &   0.958(26) & 0.936(36) &  0.896(58) & 0.87(13) & 0.62(50)  \\ 
\end{tabular}
{\caption{\footnotesize  \label{tab:0z} Same as in tab.~\ref{tab:0y} but for the ratios of $R_T(w_3)$ and computed following eq.~(\ref{eq:contT}).}}
\end{ruledtabular}
\end{table*}

\section{Summary and perspectives\label{sec:conclusions}}

In this paper we presented the results of our lattice QCD study of the exclusive semileptonic $B_s\to D_s\ell\nu_\ell$ decay form factors in the region near zero recoil (close to $q^2_{\rm max}$). The method employed here is the one proposed in ref.~\cite{blossier} that allows circumventing the problem of extrapolation in the inverse heavy quark mass and to reach the physical answer through interpolation. This is achieved by studying the successive ratios of form factors computed with heavy ``$b$"-quark mass differing by a fixed factor of $\lambda$. In that way, in the continuum limit, these ratios have a fixed value for $m_b\to \infty$ and instead of extrapolating, one interpolates to the (inverse) $b$-quark mass.

We first computed the normalization to the vector form factor relevant to the extraction of the CKM matrix element $|V_{cb}|$ from  $B(B_s\to D_s\ell\nu_\ell)$ with the light lepton in the final state $\ell \in \{e,\mu\}$. We obtained 
\bea
\gone = 1.052(46)\,,
\eea
and found that the method used here can also be employed to compute  $\gone$ for the non-strange decay modes $B(B\to D\ell\nu_\ell)$, provided the statistical sample of gauge field configurations were larger. 
We also observe that the above error bar can be significantly reduced if one imposed that the form factor ratios and the form factors themselves do not depend on the mass of the dynamical (sea) quark, which is essentially 
what we see with all of our lattice data (at all values of the lattice spacing).

Thanks to the use of twisted boundary conditions imposed on the valence charm quark, we were able to explore the region of very small momenta given to $D_s$, and therefore to compute the form factors for small recoil momenta $w\gtrsim 1$. Since we restrained our analysis to very small $w$'s, we could not estimate the accurate value of the slope of the form factor $\gw$. 

Instead, we computed the ratio of the scalar to vector form factors, $R_0(q^2)=f_0(q^2)/f_+(q^2)$, that is needed to interpret the recent discrepancy between the experimentally measured $B(B\to D\tau\nu_\tau)/B(B\to D\mu\nu_\mu)$  and its theoretical prediction within the Standard Model. Since the scalar form factor contribution to the decay rate is helicity suppressed in the Standard Model, it is much more significant for the case of the $\tau$-lepton in the final state than in the case of $\mu$. This contribution is very important in various NP scenarios. In this paper we computed $f_0(q^2)/f_+(q^2)$ by using the same method of ratios and by restraining our attention to the small recoil region. Of several small $w$'s, we quote 
\bea
\l. \frac{f_0(q^2)}{f_+(q^2)}\r|_{q^2=11.5\ \gev^2}\!\!\!\!\! = 0.77(2)\,.
\eea
Finally, in the models of physics beyond the Standard Model in which the tensor coupling to a vector boson is allowed, a third form factor might become important. Here we provide the first lattice QCD estimate of this (tensor) form factor $f_T(q^2)$ with respect to the vector one, $f_+(q^2)$. By employing the same methodology as above, in the $\msbar$ renormalization scheme and at $\mu =m_b$ we obtain
\bea
\l. \frac{f_T(q^2)}{f_+(q^2)}\r|_{q^2=11.5\ \gev^2}\!\!\!\!\! = 1.08(7)\,.
\eea 
We attempted repeating the same analysis for the case of the non-strange decay mode and found $f_0(q_0^2)/f_+(q_0^2)=0.73(4)$ for $q_0^2=11.6\ \gev^2$ (and $f_0(q_0^2)/f_+(q_0^2)=0.75(2)$ if neglecting the dependence on the sea quark mass), in agreement with the results obtained in refs.~\cite{Bailey:2012rr,nazarioA}. Finally, in the non-strange case we get $f_T(q_0^2)/f_+(q_0^2)=1.06(12)$, which becomes $1.10(7)$ if neglecting the dependence on the sea quark mass. These values show that the prospects of using this method for computing the form factors for the non-strange decay modes, $B\to D\ell\nu_\ell$, are promising provided the statistical quality of the data is improved.

\section*{Acknowledgements}
We thank  the ETMC for making their gauge field configurations publicly available. We are also grateful to B.~Blossier, V.~Lubicz and O.~P\`ene for discussions related to the subject of this paper. 
Computations of the relevant correlation functions are made on the HPC resources of IDRIS Orsay, thanks to the computing time given to us by GENCI (2013-056808). M.A. is grateful to the CNRS Liban for funding her research.

\end{document}